\documentclass[11pt]{article}
\usepackage{mydef2col}
\usepackage{markArticle}

 \title{Marketron Through the Looking Glass: From Equity Dynamics to Option Pricing in Incomplete Markets}
\shorttitle{Marketron Through the Looking Glass}
\author{
\authorstyle{Igor Halperin\textsuperscript{1}\thanks{e-mail: \url{ighalp@gmail.com}} \,
and Andrey Itkin\textsuperscript{2}\thanks{e-mail: \url{aitkin@nyu.edu}, the corresponding author}
}
\newline\newline
\textsuperscript{1}
\institution{Fidelity Investments, USA.}\\
\textsuperscript{2}
\institution{FRE Department, Tandon School of Engineering, New York University, USA.}
}

\date{}

\begin{document}

\maketitle

\centerline{\normalfont \normalsize \sffamily  \textcolor{red}{\bf ** To the memory of Tom Hurd **}}
\vspace*{0.1in}

\lettrineabstract{The Marketron model, introduced by [Halperin, Itkin, 2025], describes price formation in inelastic markets as the nonlinear diffusion of a quasiparticle (the marketron) in a multidimensional space comprising the log-price $x$, a memory variable $y$ encoding past money flows, and unobservable return predictors $z$. While the original work calibrated the model to S\&P 500 time series data, this paper extends the framework to option markets - a fundamentally distinct challenge due to market incompleteness stemming from non-tradable state variables. We develop a utility-based pricing approach that constructs a risk-adjusted measure via the dual solution of an optimal investment problem. The resulting Hamilton-Jacobi-Bellman (HJB) equation, though computationally formidable, is solved using a novel methodology enabling efficient calibration even on standard laptop hardware. Having done that, we look at the additional question to answer: whether the Marketron model, calibrated to market option prices, can simultaneously reproduce the statistical properties of the underlying asset's log-returns. We discuss our results in view of the long-standing challenge in quantitative finance of developing an unified framework capable of jointly capturing equity returns, option smile dynamics, and potentially volatility index behavior.
}

\section{Introduction} \label{Introd}

In \cite{HalperinItkin2025ar}, we presented a model of price formation in an inelastic market where dynamics are driven by both money flows and their impact on asset prices. The model treats market inflow as an investment policy of external investors. For price impact, we employ a function that captures both market inelasticity and saturation from new money (the "dumb money" effect). Since market investors' flows depend on market performance, the model creates a feedback mechanism leading to nonlinear dynamics. This results in market price dynamics being represented as nonlinear diffusion of a quasiparticle (the marketron) in a two-dimensional space defined by the log-price $x$ and a memory variable $y$. The memory variable retains information about past money flows, making the dynamics non-Markovian in log-price $x$ alone, but Markovian in the pair $(x,y)$ - similar to spiking neuron models in neuroscience.

Beyond market flows, the dynamics are influenced by return predictors, modeled as unobservable Ornstein-Uhlenbeck processes. By interpreting predictive signals as self-propulsion components of price dynamics, we treat the marketron as an active particle, allowing us to apply methods from active matter physics. Our model produces various dynamic scenarios depending on parameter choice, predicting three distinct market regimes: the "Good," the "Bad," and the "Ugly" markets. The last regime describes either a total market collapse or a corporate default event, depending on whether the model is applied to the entire market or an individual stock.

In \cite{HalperinItkin2025ar}, we calibrated the 3D version of the model to S\&P 500 return time-series using particle filtering and optimization methods. The calibrated model successfully produced a marketron potential with metastable market regimes and instanton-facilitated transitions between them. The model generates observable instantaneous defaults during simulation, with default intensities determined by model parameters. The calibrated model yields an annualized default intensity of approximately 18 basis points, aligning well with market-implied default intensities typically ranging from 10 to 50 basis points. This agreement is particularly significant as calibration used only equity market data, without incorporating credit market information.

While this calibration approach proved relatively successful, alternative market data could be used, such as S\&P500 index options or CDS market data. In this paper, we develop an approach for calibrating the model to options market data. In our framework, while stock and option prices represent tradable and liquid assets, the memory variable $y_t$ and signals $\theta_t$ are non-tradable, hidden stochastic variables. Consequently, the pricing and hedging of derivatives written on the spot price $S_t$ presents an incomplete market problem due to these hidden variables. Various methodologies have been developed to construct suitable option pricing measures in incomplete markets, e.g., see \cite{MusielaZariphopoulou2004, GrasselliHurd2007, HH2008, HendersonLiang2011, HalperinItkin2014} and references therein. In this paper,
we apply the concepts of utility-based pricing and hedging of derivatives in incomplete markets, following the approach of \cite{GrasselliHurd2007}. We derive the corresponding Hamilton-Jacobi-Bellman (HJB) equation and solve it to obtain a partial differential equation (PDE) for the option price that exhibits quadratic nonlinearity in the price and various nonlinearities in the coefficients due to the original construction of the Marketron model.

To solve this PDE, we first transform it to a nonlinear Volterra integral equation of the second kind using a generalized Duhamel's principle, as explained in \cite{ItkinMuravey2024jd, Itkin2024jd}. This approach is particularly effective given our model's assumption of constant volatilities, as the Green's function for the homogeneous problem has a known analytical form (a 3D Gaussian), making the kernel of the Volterra integral equation analytically tractable. Inspired by this form of the kernel, we use a Radial Basis Functions (RBF) method with the Gaussian kernel and quadrature rules in time to reduce our problem to solving a nonlinear (quadratic) matrix equation for the RBF coefficients. Importantly, due to our choice of the RBF kernel, all elements of these matrices can be computed in closed form since all corresponding 3D integrals can be evaluated analytically. We detail this novel result in \cref{appMatrixRBF}. To the best of our knowledge, this approach has not been previously reported in the literature.

Furthermore, we suggest a modification of this method constructed based on Strang's splitting of the nonlinear HJB PDE. We show that at every step of splitting which includes nonlinear terms, the Cole-Hopf transformation could be successfully applied, hence instead of solving a nonlinear system of equations (as at the previous step) we need to solve only three linear systems of equations where all matrices and right-hands sides can be obtained in closed form by computing the corresponding 3D integrals analytically. This modification significantly improves the performance of the pricer while providing a sufficient accuracy (second order approximation in time step). This result is also new.

Using the proposed approach, we calibrate our model to option market data and also examine the market price of risk, which is derived from the analysis of our HJB framework.

The rest of the paper is organized as follows. In \cref{model} we provide a short survey of the Marketron model as it is presented in \cite{HalperinItkin2025ar}. \cref{sect:indiff} describes the existing approaches of pricing derivatives in incomplete markets, especially the indifference pricing approach. It derives the corresponding HJB equation for the option price and shows how it can be transformed into a nonlinear PDE. In \cref{sec:Volterra} we develop a new approach for solving this PDE by first transforming it into a Volterra integral equation of the second kind. This is possible since the Green's function of the 3D PDE is known, so a generalized Duhamel's principle (see \cite{ItkinMuravey2024jd, Itkin2024jd}) can be used for doing so. In \cref{solRBF} we develop a version of the RBF method to solve this Volterra integral equation. Furthermore, in \cref{sec-solSplit} we improve this method by using Strang's splitting and the Cole-Hopf transformation to reduce the problem to solving three systems of linear equations where all system matrices can be computed in closed form since all corresponding 3D integrals can be evaluated analytically. \cref{exper} demonstrates the efficiency of this approach by describing test results for pricing SPY options within the Marketron model. In \cref{sect:calib} we calibrate the model to a daily snapshot of SPY market data and examine the market price of risk, which directly follows from our HJB framework. The final section concludes.

\section{The Marketron model} \label{model}

The Marketron model in \cite{HalperinItkin2025ar} is driven by a two-dimensional marketron potential. There we derived two Langevin equations to govern the marketron dynamics, which are determined by market flows, and also are partially driven by unobservable return predictors. While market flows are partially observable, the original model specification in \cite{HalperinItkin2025ar} treats them as another unobservable process.
In this paper, we stick to the same setting, thus treating both the market flows and return predictors as unobservable processes.
Furthermore, we used a one-dimensional Ornstein-Uhlenbeck (OU) process for a hidden signal $\theta_t$ to construct a return predictor.

The complete dynamics of the Marketron model with this specification are described by a system of stochastic differential equations (SDEs).
\begin{alignat}{2} \label{Marketron_3D}
dx_t &= \mu_x dt + \sigma d W^{(x)}_t, \qquad& \mu_x &= f(\theta_t)  + \bar{\eta} -  c(t) y_t V'_M(x_t), \\
dy_t &= \mu_y dt + \sigma_y d W{(y)}_t, \qquad&  \mu_y &= h(\theta_t) + \mu (\bar{y} - y_t) -  c(t) V_{M}(x_t), \nonumber  \\
d\theta_t &= \mu_\theta dt + \sigma_{z} d W^{(\theta)}_{t}, \qquad& \mu_\theta &= k( \hat{\theta} - \theta_t), \nonumber \\
x(0) &= x, \quad y(0) = y, \quad \theta(0) = \theta. \nonumber
\end{alignat}
Here, $t$ represents time, and $W^{(x)}_t$, $W^{(y)}_t$, $W^{(\theta)}_t$ are independent Brownian motions (BM). The term $c(t) \geq 0$ is a deterministic function of time, while $\sigma$, $\sigma_y$, $\sigma_z$ represent volatilities of the corresponding stochastic processes. The parameters $\bar{\eta}$, $\mu$, $\bar{y}$, $k$, $\hat{\theta}$ are to be determined through model calibration to market data.
The functions $f$ and $h$ are functions of bounded variation, introduced to control the amplitude of the random signal $\theta_t$. In \cite{HalperinItkin2025ar}, they were chosen as
\begin{align} \label{fhparam}
f(\theta) = a_1(t)/\left(1 + e^{-b_1 \theta} \right), \qquad h(\theta) = a_2(t)/\left(1 + e^{-b_2 \theta} \right), \\
a_1(t) = k_{1,x} \cos(k_{2,x} + k_{3,x} t), \qquad a_2(t) = k_{1,y} \sin(k_{2,y} + k_{3,y} t), \nonumber
\end{align}
\noindent where $b_1, b_2, k_{i,x}, k_{i,y}, , i \in [1,3]$ are constants determined by calibration. Also, in \cite{HalperinItkin2025ar} we assumed that  $ c(t) $ does not change with time, i.e. $c(t) = c$.
Finally, we adopt a two-dimensional (2D) formulation of the Langevin dynamics in
\eqref{Marketron_3D}, where the drifts $ \mu_x, \mu_y $ are interpreted
as negative gradients of a 2D potential $ V(x,y) $ (the marketron potential) which has the following form\footnote{When calibrating this model to market option prices, we will slightly modify the definition of $V_M(x)$ to make our calibration procedure more tractable, see \cref{appLinear}.}
\begin{align} \label{Langevin_potential_exact}
V(x, y) &= -  \bar{\eta} x   + c(t) y V_{M}(x) + \frac{1}{2} \mu \left(y - \bar{y} \right)^2, \\
V_{M}(x) &= \frac{1}{\bar{\epsilon}} \left[ (\bar{\epsilon} -1) \left(e^{-x}  -1 \right)  + \frac{1}{\kappa}  \log \frac{1 + \kappa e^{-x}}{1 + \kappa} \right], \quad V_M'(x) = - e^{-x} \left( 1 - \frac{g}{e^{x} + \varepsilon g} \right), \quad \kappa = g \bar{\epsilon}, \nonumber
\end{align}
Parameters $g$ and $\bar{\epsilon}$ in \eqref{Langevin_potential_exact} are introduced to serve two different objectives. While the coupling constant parameter $g$ "couples" the investors policy and the price, the second parameter $\bar{\epsilon}$ serves as a {\it regularization parameter}: it controls the behavior of the policy at large negative $x_t$.

\section{Risk-minimal measure and option pricing} \label{sect:indiff}

The primary objective of this paper is to analyze the Marketron model governed by \eqref{Marketron_3D}, by calibrating it to market data from derivatives (options) written on the underlying stock $S_t$, where $x_t = \log (S_t/S_*)$ and $S_*$ is a constant. This analysis requires a fast and reliable option pricing method to determine option values under the risk-neutral (or risk-minimal) measure. In this framework, while stock and option prices represent tradable and liquid assets, the memory variable $y_t$ and signals $\theta_t$ are non-tradable, hidden stochastic variables. Consequently, the pricing and hedging of derivatives written on the spot price $S_t$ presents an incomplete market problem due to these hidden variables.

In complete markets, it is well-established that the equivalent martingale measure is unique, which uniquely determines option prices (as exemplified in the Black-Scholes model). However, in incomplete markets, there exist infinitely many equivalent martingale measures, necessitating the selection of an appropriate measure from this set to determine option prices. Various methodologies have been developed to construct suitable option pricing measures in incomplete markets, e.g., see \cite{MusielaZariphopoulou2004, GrasselliHurd2007, HH2008, HendersonLiang2011, HalperinItkin2014} and references therein. In this paper, we apply the concepts of utility-based pricing and hedging of derivatives in incomplete markets, following the approach of \cite{GrasselliHurd2007}.

Accordingly, we seek an optimal hedging portfolio, which corresponds to the strategy of an investor with initial wealth $z$ who faces a (discounted) financial liability (the option) $\calV$ maturing at time $T$. Let $U: \mathbb{R} \rightarrow \mathbb{R} \cup\{-\infty\}$ be a utility function, assumed to be a concave, strictly increasing and differentiable function. The investor aims to solve the stochastic control problem
\begin{equation} \label{optHedging}
u(z) = \sup _{\mathcal{A}} \EE\left[U\left(Z_T - \calV \right) \mid Z_0 = z \right],
\end{equation}
\noindent where $Z_T$ is the discounted terminal wealth obtained when investing $h_t S_t$ dollars in the risky asset and $\rho_t C_t$ dollars in a riskless cash account. The cash account value $C_t$, initialized at $C_0 = 1$, follows
\begin{equation}
d C_t = r_t C_t d t.
\end{equation}
For simplicity, we assume a constant interest rate, i.e., $r_t = r$.

The domain of optimization $\mathcal{A}$ in \eqref{optHedging} is restricted to self-financing portfolios, that is, to wealth processes satisfying
\begin{equation} \label{selfFin}
C_t Z_t := h_t S_t + \rho_t C_t = z + \int_0^t h_u d S_u + \int_0^t \rho_u d C_u.
\end{equation}
The option price $\calV$ is assumed to be a random variable of the form $\calV_T = \calV(S_T, y_T, \theta_T)$, where $\calV: \mathbb{R}_+ \times \mathbb{R}_+ \times \mathbb{R}_+ \to \mathbb{R}$ is a Borel-measurable function.

Introducing the discount price $s_t = S_t/C_t$ and applying \Ito lemma to \eqref{Marketron_3D}, we obtain
\begin{alignat}{2} \label{Marketron_3D-1}
ds_t &= s_t \left[ \bar{\mu}_x dt + \sigma d W_t \right], &\qquad \bar{\mu}_x  &= \mu_x + \frac{1}{2} \sigma^2  - r, \\
dy_t &= \mu_y dt + \sigma_y d W^y_t, &\qquad
d\theta_t &= \mu_\theta dt + \sigma_{\theta} d W^\theta_t. \nonumber
\end{alignat}
The self-financing condition \eqref{selfFin} implies that the discounted wealth process satisfies
\begin{equation} \label{sdeZ}
d Z_t = h_t d s_t  = h_t s_t \left[ \bar{\mu}_x dt + \sigma d W_t \right],
\end{equation}
\noindent while the holdings in the cash account are given by $\rho_t = Z_t - h_t s_t$.

Given the Markovian nature of our problem in equations \cref{Marketron_3D,Marketron_3D-1}, the optimal hedging problem \eqref{optHedging} can be reformulated as
\begin{equation} \label{utilFunc}
u(t, z, s, y, \theta) = \sup_{h \in \mathcal{A}_t} \EE_{t,s,y,\theta} \left[U \left(Z_T - \calV(s_T, y_T, \theta_T) \right) \mid Z_t = z \right], \qquad t \in (0,T),
\end{equation}
\noindent where $z \in \mathbb{R}$ represents an arbitrary level of wealth, $\mathcal{A}_t$ denotes the set of admissible portfolios starting at time $t$, and $\EE_{\tau,s,y,\theta}$ denotes expectation with respect to the joint probability law at time $\tau$ of the processes $s_u, y_u, \theta_u$ satisfying \cref{Marketron_3D,Marketron_3D-1} for $u \ge \tau$ with the initial conditions $s_\tau = s, y_\tau = y, \theta_\tau = \theta$.

In this paper we employ an exponential utility of the form
\begin{equation} \label{expUtil}
U(z) = - e^{-\gamma z},
\end{equation}
\noindent where $\gamma > 0$ represents the risk-aversion parameter. The exponential utility function is particularly useful because it allows us to factorize the value function $u(t, z, s, y, \theta)$ in \eqref{utilFunc} as
\begin{align} \label{utilFunc2}
u(t, z, s, y, \theta) &= \sup_{h \in \mathcal{A}_t} \EE_{t,s,y,\theta} \left[ - e^{-\gamma \left[ z + \Psi(t,T) - \calV(s_T, y_T, \theta_T) \right] } \right] \\
&= - e^{-\gamma z} \inf_{h \in \mathcal{A}_t} \EE_{t,s,y,\theta} \left[ - e^{-\gamma \left[\Psi(t,T) - \calV(s_T, y_T, \theta_T) \right]} \right] =
U(z) v(t,x,y,\theta), \qquad \Psi(t,T) = \int_t^T h_u ds_u. \nonumber
\end{align}

Finally, we need to define the certainty equivalent, \cite{GrasselliHurd2007} for the claim $\calV$ at time $t$ as the process $\calC = \calC(t,z,s,y, \theta)$ satisfying the equation
\begin{equation} \label{CertEq}
U(z - \calC) = \EE_{t,s,y,\theta}\left[ U\left( z + \Psi(t,T) - \calV(s_T, y_T, \theta_T) \right) \right].
\end{equation}
It represents the amount which, when subtracted from wealth $z$ at time $t$, yields the same deterministic utility value as the optimal expected utility of terminal wealth. This calculation starts with initial wealth $z$ and accounts for claim $\calV$ at terminal time $T$.

An agent with utility $U$ and wealth $z$ at time $t \in(0, T)$ will charge a premium for issuing a liability $\calV$ maturing at $T$. The indifference price for the claim $\calV$ is defined to be the premium that makes the agent indifferent between making the deal or not, that is, the unique solution $\pi^\calV=\pi^\calV(t, z, s, y, \theta)$ (if it exists) to the equation
\begin{equation}
\sup _{h \in \mathcal{A}_t} \EE_{t, s, y, \theta} \left[ U\left( z + \Psi(t,T) \right) \right] = \sup_{h \in \mathcal{A}_t}
\EE_{t, s, y, \theta} \left[ U \left( z + \pi^\calV + \Psi(t,T) - \calV(s_T, y_T, \theta_T) \right) \right].
\end{equation}
From the definition of the certainty equivalent $\calC$, it is seen that this equation is equivalent to
\begin{equation}
U \left(z - \calC^0 \right) = U \left(z + \pi^\calV - \calC \right),
\end{equation}
\noindent hence the indifference price is given by
\begin{equation} \label{indifPrice}
\pi^\calV(t, z, s, y, \theta) = \calC \left(t, z + \pi^\calV(t, z, s, y, \theta), s, y, \theta \right) - \calC^0(t, z, s, y, \theta).
\end{equation}

It follows from the definition of the exponential utility in \eqref{utilFunc} and \eqref{CertEq} that
\begin{equation} \label{defC}
\calC(t,s,y,\theta) = \frac{1}{\gamma} \log v(t, s, y, \theta).
\end{equation}
Also, by setting $\calV=0$, we obtain
\begin{equation}
\calC^0(t,s,y,\theta) = \frac{1}{\gamma} \log v^0(t, s, y, \theta), \qquad v^0(t, s, y, \theta) = \inf_{h \in \mathcal{A}_t} \EE_{t,s,y,\theta} \left[ - e^{-\gamma \Psi(t,T)} \right].
\end{equation}
Combining these equations with \eqref{indifPrice} yields
\begin{equation} \label{indifPrice2}
\pi^\calV(t, s, y, \theta) = \calC(t, s, y, \theta) - \calC^0(t, s, y, \theta) = \frac{1}{\gamma} \log \frac{v(t, s, y, \theta)}{v^0(t, s, y, \theta)}.
\end{equation}

\subsection{Partial Differential Equation (PDE) for the Indifference Option Price}

By the dynamic programming principle, the value function $u(t, z, s, y, \theta)$ defined in \eqref{utilFunc2}, satisfies the Hamilton-Jacobi-Bellman equation (see \eqref{Marketron_3D,Marketron_3D-1,sdeZ})
\begin{align} \label{HJB}
0 = \fp{u}{t} &+ \frac{1}{2}\sigma^2 s^2 \sop{u}{s} + \frac{1}{2}\sigma_y^2 \sop{u}{y} + \frac{1}{2}\sigma_\theta^2 \sop{u}{\theta} + s \bar{\mu}_x \fp{u}{s} + \mu_y \fp{u}{y} + \mu_\theta \fp{u}{\theta} \\
&+ \max_{h} \left[ \frac{1}{2} h^2 s^2 \sigma^2 \sop{u}{z} +  h s^2 \sigma^2 \cp{u}{z}{s} + h s \bar{\mu}_x \fp{u}{z}  \right], \nonumber
\end{align}
\noindent subject to the terminal condition
\begin{equation} \label{tc}
u(T, z, s, y, \theta) = - e^{-\gamma\left( z - \calV(s_T, y_T, \theta_T) \right) }.
\end{equation}

The solution of the maximization part of \eqref{HJB} is given by
\begin{equation} \label{hSol}
h^\calV(t, s, y, \theta) = \frac{1}{\gamma} \frac{\partial_s v}{v} + \frac{\bar{\mu}_x}{\gamma s \sigma^2}.
\end{equation}
Consequently, the value function $v(t,s,y,\theta)$ satisfies the following PDE
\begin{align} \label{PDE}
\fp{v}{t} &+ \frac{1}{2}\sigma^2 s^2 \sop{v}{s} + \frac{1}{2}\sigma_y^2 \sop{v}{y} + \frac{1}{2}\sigma_\theta^2 \sop{v}{\theta} + \mu_y \fp{v}{y} + \mu_\theta \fp{v}{\theta} -\frac{1}{2} \sigma^2 s^2 \frac{v^2_s(t,s,y,\theta)}{v(t,s,y,\theta)} - \frac{\bar{\mu}^2_x}{2 \sigma^2} v(t,s,y,\theta) = 0,
\end{align}
\noindent subject to the terminal condition
\begin{equation} \label{tc1}
v(T, s, y, \theta) = e^{\gamma \calV(s_T, y_T, \theta_T)}.
\end{equation}

The same PDE must be used to find $v^0(t,s,y,\theta)$, but with the terminal condition
\begin{equation} \label{tc0}
v^0(T, s, y, \theta) = 1.
\end{equation}

Taking into account the definition in \eqref{defC}, applying the change of variables $s \to x$ and also discounting the certainty equivalent price, it follows that $\calC(t,x,y,\theta)$ solves the PDE
\begin{align} \label{PDEc}
\fp{\calC}{t} &+ \frac{1}{2}\sigma^2 \sop{\calC}{x} + \frac{1}{2}\sigma_y^2 \sop{\calC}{y} + \frac{1}{2}\sigma_\theta^2 \sop{\calC}{\theta} + \bar{\eta} \fp{\calC}{x} + \mu_y \fp{\calC}{y} + \mu_\theta \fp{\calC}{\theta}  - \frac{\bar{\mu}^2_x}{2 \gamma \sigma^2} - r \calC \\
&+ \frac{1}{2} \gamma \sigma_y^2 \left( \fp{\calC}{y} \right)^2 + \frac{1}{2} \gamma \sigma_\theta^2 \left( \fp{\calC}{\theta} \right)^2 = 0, \nonumber
\end{align}
\noindent subject to the terminal condition
\begin{equation} \label{tcC}
\calC(T, x, y, \theta) = \calV(x_T, y_T, \theta_T).
\end{equation}

The boundary conditions and option payoff depend on the type of the option.  For Call options, they are
\begin{align} \label{callBC}
\calV(x_T, y_T, \theta_T) &= (S_* e^{x_T} - K)^+, \qquad \calC(t, x \downarrow -\infty, y, \theta) = 0, \qquad \calC(t, x \uparrow \infty, y, \theta) = S_* e^x,
\end{align}
\noindent and for Put options they are
\begin{align} \label{putBC}
\calV(x_T, y_T, \theta_T) &= (K - S_* e^{x_T})^+, \qquad \calC(t, x \uparrow \infty, y, \theta) = 0, \qquad \calC(t, x \downarrow -\infty, y, \theta) = K.
\end{align}

Additionally, solving \eqref{PDEc} with the terminal condition
\begin{equation} \label{tc00}
\calC(T, x, y, \theta) = 0,
\end{equation}
\noindent yields $\calC^0(t,x,y,\theta)$. The indifference option price $\pi^\calV(t, x, y, \theta)$ can then be obtained from \eqref{indifPrice2}.

\subsection{Market price of risk} \label{mpr}

As shown in \cite{GrasselliHurd2007}, the utility based price of risk is a vector process, which in our case has three components corresponding to the number of stochastic factors. However, as mentioned in that paper, both the optimal martingale measure $Q^\calV$ and the utility based price of risk $\lambda_t^\calV$ are specifically related to the claim $\calV$, and therefore, do not constitute a direct generalization of the paradigm of pricing by expectation with respect to a risk adjusted measure valid for all claims. For instance, the indifference price $\pi^\calV$ is not linear in the claim $\calV$, with the obvious effect that $Q^\calV$ fails to be a pricing measure even for multiples of $\calV$, let alone for other unrelated claims.

However, the Davis price can be calculated as an expectation with respect to the risk adjusted measure $Q$ obtained as the dual solution for the optimal investment problem according to the utility function $U$. This measure induces a utility based market price of risk valid for all claims, which in the case of an exponential utility is obtained from $\mathcal{C}^0$.

For the exponential utility function all components of the utility based market price of risk can be obtained in closed form. Indeed,
based on \cite{GrasselliHurd2007}, the optimal martingale measure $Q^\calV$ and the optimal wealth $Z^\calV_T = z + (h^\calV \cdot S)_T$
are related by the fundamental equation (see \eqref{utilFunc}
\begin{equation}
U'(Z^\calV_T - \calV) = \xi \frac{d Q^\calV}{dP}, \qquad \xi = u'(z).
\end{equation}
Introducing the density process for the measure $Q^\calV$ as
\begin{equation}
\Lambda^\calV_t = \EE_t \left[ \frac{d Q^\calV}{dP} \right] = \frac{1}{\xi} \EE_t \left[ U'(Z^\calV_T - \calV) \right],
\end{equation}
\cite{GrasselliHurd2007} define the utility based price of risk associated with the claim $\calV$ as the vector process $\lambda_t^\calV$, which in our model and notation reads
\begin{equation}
\lambda_t^\calV = \left( \lambda^{(x)}_t, \lambda^{(y)}_t, \lambda^{(\theta)}_t \right).
\end{equation}
This process satisfies the equation
\begin{equation}
\frac{d \Lambda_t^\calV}{\Lambda_t^\calV} = - \left[ \lambda^{(x)}_t d W^{(x)}_t  + \lambda^{(y)}_t d W^{(y)}_t
+ \lambda^{(\theta)}_t d W^{(\theta)}_t \right].
\end{equation}

For the case of exponential utility this reduces to (see \eqref{CertEq})
\begin{align} \label{densityEq}
\Lambda_t^\calV &= - \frac{\gamma}{\xi} \EE_{t,s,y,\theta}\left[ U\left( Z_T^\calV + \Psi(t,T) - \calV(s_T, y_T, \theta_T) \right) \right] =
- \frac{\gamma}{\xi} U\left( Z_t^\calV - \calC^\calV_t \right) = \frac{ e^{-\gamma\left(Z_t^\calV - \calC^\calV_t\right)} } {e^{-\gamma\left(z - \calC_0^\calV\right)}},
\end{align}
\noindent due to the definition of the certainty equivalence process and the relationship
\begin{equation}
\xi = u'(z) = U'(z - \calC_0^\calV).
\end{equation}
Applying \Ito's lemma to the martingale $\Lambda^\calV_t$, comparing with \eqref{densityEq} and taking into account that all stochastic factors in \eqref{Marketron_3D} are uncorrelated, the most important component $\lambda^\calV_{t,x}$ (since other stochastic variables $y, \theta$ are unobservable) we obtain
\begin{equation} \label{mprX}
\lambda^{(x)}_t = \gamma[ h_t^\calV - \partial_s \calC^\calV) \sigma s.
\end{equation}
Finally, recall the solution of the HJB equation in \eqref{hSol}
\begin{equation*}
h^\calV(t, s, y, \theta) = \frac{1}{\gamma} \frac{\partial_s v}{v} + \frac{\bar{\mu}_x}{\gamma s \sigma^2}.
\end{equation*}
Together with \eqref{mprX} and \eqref{defC} this yields
\begin{equation}
\lambda^{(x)}_t = \frac{\mu_x - r}{\sigma},
\end{equation}
\noindent and thus, it does not depend on the claim $\calV$. In other words, the component related to the first Brownian motion $W^{(x)}_t$ is formally identical to the market price of risk of a complete market. However, due to the definition of $\mu_x$ in \eqref{Marketron_3D}, it does depend on the parameters of the Marketron potential.  More precisely, the nonlinear drift introduced by the Marketron model causes the market price of risk to depend on the parameters of the Marketron potential, as well as on the initial values of the factors $x, y, \theta$ (states). In particular, if $x \to -\infty$ in follows from \cref{Marketron_3D,Langevin_potential_exact} that $\mu_x \to -\infty$. The dependence of the market price of risk on states has been already investigated in the literature, see e.g., \cite{DaiSingleton:99,Sbuelz2012,Mijatovic2013} among others.

\subsection{Transformation of \eqref{PDEc} into a Volterra integral equation} \label{sec:Volterra}

As follows from the definitions of drifts in \eqref{Marketron_3D}, we have $\mu_x = \mu_x(t, x, y, \theta), \mu_y = \mu_y(t, x, y, \theta), \mu_\theta = \mu_\theta(t, \theta)$. Consequently, \eqref{PDEc} represents a nonlinear PDE in the dependent variable $\calC$ where the coefficients are nonlinear functions of $x,\theta$ and linear in $y$. In general, this equation does not admit an analytical solution and requires numerical method.  However, using a numerical option pricer for calibrating the Marketron model to market option quotes would be computationally inefficient, particularly given the high dimensionality of the calibration problem (eighteen parameters). Therefore, an alternative approach is needed, which we describe below.

The main idea of such an approach consists in splitting the solution into two steps. First, we solve a "homogeneous" PDE containing only the highest-order derivatives of $\mathcal{C}$ with respect to $x$, $y$, and $\theta$ using the Green's function method. Second, we apply a generalized Duhamel's principle, \cite{ItkinMuravey2024jd, Itkin2024jd}, to transform \eqref{PDEc} into a nonlinear Volterra integral equation of the second kind for $\calC$, which also solves the complete PDE in \eqref{PDEc}. This approach is particularly effective given our model 's assumption of constant volatilities, as the Green's function for the homogeneous problem has a known analytical form, making the kernel of the Volterra integral equation analytically tractable. Consequently, instead of solving the full problem in \eqref{PDEc} numerically, we first partially solve it analytically, and then apply numerical procedures only to a simplified equation.

To proceed, technically we represent \eqref{PDEc} in the form
\begin{gather} \label{SemiLinearPDE}
\fp{\calC}{t} + \frac{1}{2}\sigma^2 \sop{\calC}{x} + \frac{1}{2}\sigma_y^2 \sop{\calC}{y} + \frac{1}{2}\sigma_\theta^2 \sop{\calC}{\theta} +
\Phi = 0, \qquad (t,x,y,\theta) \in [0,T) \times \mathbb{R}^3, \\
\Phi = \Phi(\calC_x, \calC_y, \calC_\theta, t, x, y, \theta) = \bar{\eta} \fp{\calC}{x} + \mu_y \fp{\calC}{y} + \mu_\theta \fp{\calC}{\theta} + \frac{1}{2} \gamma \sigma_y^2 \left( \fp{\calC}{y} \right)^2 + \frac{1}{2} \gamma \sigma_\theta^2 \left( \fp{\calC}{\theta} \right)^2 - r \calC - \frac{\bar{\mu}^2_x}{2 \gamma \sigma^2}. \nonumber
\end{gather}

Using the approach of \cite{Itkin2024jd,Hunter2014} (see also \cite{BenArtzi2007}), \eqref{SemiLinearPDE} can be interpreted as a parabolic equation with nonlinear gradient terms
\begin{equation} \label{ode2}
u_t = \Xi u + g(u, u_x)
\end{equation}
\noindent where the linear operator $\Xi$ generates a semigroup on a Banach space $X$, and
\begin{equation*}
g: \mathcal{D}(F) \subset X \rightarrow X
\end{equation*}
\noindent is a nonlinear function. For semi-linear PDEs, $g = g(u)$ typically depends on $u$ and not on its spatial derivatives.
In such a case, \eqref{ode2} represents a linear PDE perturbed by a zeroth-order nonlinear term. However, our case falls under the category of  viscous Hamilton-Jacobi equations, \cite{BenArtzi2007}, where $g = g(u, u_x)$ depends on both $u$ and its spatial derivatives.

If $\{\mathrm{T}(t)\}$ denotes the semigroup generated by $\Xi$, we can reformulate \eqref{ode2} as an integral equation for $u: [0, T] \rightarrow X$
\begin{equation} \label{odeSol2}
u(t)=\mathrm{T}(t) u(0)+\int_0^t \mathrm{~T}(t-s) g(u(s), u_x(s)) ds.
\end{equation}
If solutions to this integral equation exist and possess sufficient regularity, they also satisfy \eqref{ode2}.

Applying this approach to \eqref{SemiLinearPDE}, we find that the operator $A$ takes the form
\begin{equation}
\Xi = \frac{1}{2}\sigma^2 \sop{}{x} + \frac{1}{2}\sigma_y^2 \sop{}{y} + \frac{1}{2}\sigma_\theta^2 \sop{}{\theta}.
\end{equation}
The Green's function for the homogeneous PDE in \eqref{SemiLinearPDE} is given by, \cite{Polyanin2002}
\begin{align} \label{Green}
G(\tau; x,y,\theta|\xi, \eta, \zeta) = \frac{1}{(2 \pi \tau)^{3/2} \sigma \sigma_y \sigma_\theta} \exp \left[ - \frac{(x-\xi)^2}{2 \sigma^2 \tau} - \frac{(y-\eta)^2}{2 \sigma_y^2 \tau} - \frac{(\theta-\zeta)^2}{2 \sigma_\theta^2 \tau} \right], \qquad \tau = T - t,
\end{align}
\noindent and
\begin{equation}
G(0; x,y,\theta|\xi, \eta, \zeta) = \delta(x-\xi) \delta(y-\eta) \delta(\theta - \zeta).
\end{equation}

Accordingly, from \eqref{odeSol2} we derive
\begin{align} \label{Volterra}
\calC(\tau,x,y,\theta) &= \calI(\tau,x) + \calJ(\tau, x,y,\theta; \calC), \\
\calI(\tau, x) &= \int_{-\infty}^\infty \int_{-\infty}^\infty \int_{-\infty}^\infty G(\tau; x,y,\theta|\xi, \eta, \zeta) \calC(T-\tau,\bm{p}) d\bm{p}, \nonumber \\
\calJ(\tau, x,y,\theta; \calC) &= \int_0^{\tau(t)} \int_{-\infty}^\infty \int_{-\infty}^\infty \int_{-\infty}^\infty \bar{\Phi}(\nu, \bm{p}; \calC) G(\tau-\nu; x,y,\theta| \bm{p}) d\nu d\bm{p}, \qquad \bm{p} = \Big\{ \xi, \eta, \zeta \Big\}, \quad d\bm{p} = d\xi d\eta d\zeta, \nonumber \\
\bar{\Phi}(\nu, \bm{p}; \calC) &= \bar{\eta} \calC_\xi(\nu, \bm{p}) + \mu_y(\nu, \bm{p}) \calC_\eta(\nu, \bm{p}) + \mu_\theta(\zeta) \calC_\zeta(\nu, \bm{p}) + \frac{1}{2} \gamma \sigma_y^2 \calC^2_\eta(\nu, \bm{p}) + \frac{1}{2} \gamma \sigma_\theta^2 \calC^2_\zeta(\nu, \bm{p}) - r \calC - \frac{\bar{\mu}^2_x(\nu, \bm{p})}{2 \gamma \sigma^2}. \nonumber
\end{align}
This equation is a nonlinear Volterra integral equation of the second kind with respect to $\calC$.

Due to the explicit representation of the Green's function in \eqref{Green} and the payoff functions in  \cref{callBC,putBC}, the first integral $\calI(\tau, x)$ in \eqref{Volterra} can be computed in closed form. For Call options this yields
\begin{align} \label{callI1}
\calI(\tau,x) &= S_* e^{\frac{\sigma^2 \tau }{2} + x} \Psi\left( \frac{x +\log \left(\frac{S_*}{K}\right)}{\sigma \sqrt{\tau }} + \sigma \sqrt{\tau} \right) - K \Psi\left(\frac{x + \log \left(\frac{S_*}{K}\right)}{\sigma \sqrt{\tau }}\right),
\end{align}
\noindent while for Put options this yields
\begin{align} \label{putI1}
\calI(\tau,x) &= - S_* e^{\frac{\sigma^2 \tau }{2} + x} \Psi\left( - \frac{x + \log \left(\frac{S_*}{K}\right)}{\sigma \sqrt{\tau }} - \sigma \sqrt{\tau} \right) + K \Psi\left(- \frac{x + \log \left(\frac{S_*}{K}\right)}{\sigma \sqrt{\tau }}\right),
\end{align}
\noindent where $\Psi(x)$ is the normal CDF function, \cite{as64}.

\subsection{Solving the Volterra Integral equation using a Radial Basis Function (RBF) method} \label{solRBF}

Although we obtained $I_0$ in closed form, solving the full Volterra integral equation in \eqref{Volterra} requires numerical methods. While various numerical approaches exist in the literature, we follow \cite{CarrItkinMuraveyHeston} and employ the Radial Basis Functions (RBF) method. This choice is particularly advantageous because the integral kernel (the Green's function) in \eqref{Volterra} is Gaussian, allowing us to compute certain integrals in the right-hand side of \eqref{Volterra} analytically when using Gaussian RBF.

RBF interpolation has proven highly effective for problems of intermediate dimensionality ($10 > d > 3$), including applications in mathematical finance, \cite{YCHon3, Fasshauer2, Pettersson, FornbergFlyer2015}; see also references in \citep{Assari2019}. The method's key advantages include exponential convergence with increasing node count and its meshless nature, enabling high-resolution solutions with relatively few discretization points.

To ensure a clear exposition, we begin with definitions following \citep{Assari2019,ItkinMuraveySabrJD}. A function $\Theta: \mathbb{R}^{d} \rightarrow \mathbb{R}$ is called radial if there exists a univariate function $\phi: [0, \infty) \rightarrow \mathbb{R}$ such that
\begin{equation}
\Theta(\bm{w}) = \phi(r),
\end{equation}
\noindent where $r=|\bm{w}|$ and $|\cdot|$ denotes a norm in $\mathbb{R}^{d}$. In this paper, we specifically consider the Euclidean norm. Let $W = \left\{\bm{w}_{1},\ldots,\bm{w}_{N}\right\}$ be a set of scattered points in the domain $\Omega \subset \mathbb{R}^{d}$. A function $u(\bm{w})$ at an arbitrary point $\bm{w} \in \Omega$ can be approximated using the global radial function $\phi(|\bm{w}|)$ through a linear combination
\begin{equation} \label{GlInterp}
u(\bm{w}) \approx \mathcal{G}_{N} u(\bm{w}) = \sum_{i=1}^{N} c_{i} \phi\left(\left\|\bm{w}-\bm{w}_{i}\right\|\right), \quad \bm{w} \in \Omega,
\end{equation}
\noindent where the coefficients $\left\{c_{1}, \ldots, c_{N}\right\}$ are determined by the interpolation conditions
\begin{equation} \label{VoltRBF}
\mathcal{G}_{N} u\left(\bm{w}_{i}\right)=u\left(\bm{w}_{i}\right)=u_{i}, \quad i=1, \ldots, N.
\end{equation}

In the literature various choices of the RBFs are available. A notable example is the Gaussian RBF
\begin{equation} \label{gauss}
\Theta(\bm{w}) = e^{-\varepsilon \|\bm{w}\|^{2}},
\end{equation}
\noindent where $\varepsilon > 0$ is the shape parameter. This function is strictly positive-definite in $\mathbb{R}^{d}$, ensuring that the expansion in \eqref{GlInterp} is non-singular.

RBF methods are meshfree, operating without requiring regular grids in $\tau$ and $\bm{p}$ (unlike finite difference methods). Consider a 4D set of collocation nodes ${ \tau_i, x_{k(i)}, y_{j(i)}, \theta_{l(i)} }$ where $i = 1,\ldots,N$, $k(i)=1,\ldots,N_{i,k}$, $j(i)=1,\ldots,N_{i,j}$, $l(i)=1,\ldots,N_{i,l}$. Substituting these nodes into \eqref{GlInterp} and then plugging the RBF approximation into the Volterra integral equation yields a system of equations for coefficients $c_{i,j,k,l}$. For example, let us use a regular temporal grid $\Gamma_i$ with $N_i$ nodes $\Gamma_i: \tau_{1,i},\ldots,\tau_{N_i}$, so we have $N_{i,k}$ nodes in $x$, $N_{i,j}$ nodes in $y$ and $N_{i,l}$ nodes in $\theta$, respectively for each $i$. Even for linear Volterra equations, the resulting coefficient matrix is dense, leading to $O((N_k N_j N_l)^3)$ computational complexity when using direct solvers. While this cost is prohibitive for practical applications, iterative solvers with suitable preconditioners can improve efficiency.

Also, various methods exist to reduce the dimensionality of this problem, including local RBFs, RBFs with an improved basis, etc., (see \cite{CarrItkinMuraveyHeston} for a brief survey and additional references). In this paper, however, we employ a simple version of the global Gaussian RBF method for two reasons. First, we aim to demonstrate that the proposed method, even when combined with Gaussian RBFs, provides reasonable option prices. Second, this global method achieves faster computation times compared to the FD method. Further refinements of the method could be explored in future research.

Fortunately, this system can be significantly simplified through the following observation. Since the variables $y_t, \theta_t$ and their initial values $y, \theta$ are unobservable, we do not need to compute the option value for all possible values of $y$ and $\theta$. While we could treat these initial values as calibration parameters, our model already contains eighteen such parameters, and adding more could lead to overfitting. Therefore, following \cite{HalperinItkin2025ar}, we set $\theta = 0$ while the initial value $y(0)$ either remains a calibration parameter, or is also set to zero.

Further for simplicity, let us introduce a uniform grid $\Gamma: 0 = \tau_0,\tau_1,\ldots,\tau_N = \tau(0), \, \tau_i - \tau_{i-1} = \Delta \tau = \tau(0)/(N+1) \,\, \forall i \in [1,N]$ in time $\tau$, so, this grid contains $N+1$ nodes. Let us solve \eqref{Volterra} sequentially in time on this grid. The temporal integral at time $\tau_i$ in the definition of $\calJ(\tau, x,y,\theta; \calC)$ in \eqref{Volterra} can be approximated using some quadratures, e.g., the trapezoidal rule\footnote{This can be easily relaxed by using higher order quadrature rules.}, \cite{davis2014methods}
\begin{align} \label{trapezoid}
\calJ(\tau_i, x,y,\theta; \calC) &= \calJ(\tau_{i-1}, x,y,\theta; \calC) + \frac{1}{2} \Delta \tau  \left[ \calK(\Delta \tau_i, x, y, \theta; \calC) + \calK(0, x, y, \theta; \calC) \right], \\
\calK(\Delta \tau_i, x, y, \theta; \calC) &= \int_{-\infty}^\infty \int_{-\infty}^\infty \int_{-\infty}^\infty \bar{\Phi}(\tau_{i-1}, \bm{p}; \calC)\, G(\Delta \tau_i; x,y,\theta| \bm{p}) d\nu d\bm{p}, \qquad \calJ(0, x, y, \theta; \calC) = 0, \nonumber \\
\calK(0, x, y, \theta; \calC) &= \int_{-\infty}^\infty \int_{-\infty}^\infty \int_{-\infty}^\infty \bar{\Phi}(\tau_i, \bm{p}; \calC)\, G(0; x,y,\theta| \bm{p}) d\nu d\bm{p} = \bar{\Phi}(\tau_i, x, y, \theta; \calC). \nonumber
\end{align}
Thus, at every time $\tau_i$ to find the unknown values of $\calC(\tau_i, x,y,\theta)$ one needs to solve the nonlinear equation
\begin{align} \label{vRBF2}
\calC(\tau_i,x,y,\theta) &- \frac{1}{2} \Delta \tau_i \bar{\Phi}(\tau_i, x, y, \theta; \calC) = \calG(\tau_i,x, y, \theta), \\
 \calG(\tau_i,x, y, \theta) &= \calI(\tau_i,x) + \calJ(\tau_{i-1},x, y, \theta; \calC) + \frac{1}{2} \Delta \tau \calK(\Delta \tau_i, x, y, \theta; \calC) \nonumber \\
 &= \calI(\tau_i,x) + \Delta \tau \sum_{m=1}^{i} \calK(\Delta \tau_m, x, y, \theta; \calC). \nonumber
\end{align}
Bear in mind that in the expressions for $\calK$ and $\calJ$ at time $\tau_{i-1}$, the RBF coefficients $c_{i-1,k,j,l}$ are already known from the previous time step.

By substituting $\bar{\Phi}$ from \eqref{Volterra}, $\calC$ from \eqref{GlInterp}, and $\Theta$ from \eqref{gauss} into \eqref{vRBF2}, we obtain the following system of equation
\begin{align} \label{VolterraRBF}
\sum_{k,j,l} c_{i,kjl} & \Big\{ e^{-\varepsilon \left[(x-x_k)^2 + (y-y_j)^2 + (\theta-\theta_l)^2 \right]} - \frac{1}{2} \Delta \tau_i \bar{\Psi}_1(\tau_i, x, y, \theta, x_k, y_j, \theta_l) \Big\} \\
&- \sum_{k,j,l} \sum_{k^*,j^*,l^*} c_{i,kjl} c_{i,k^*j^*l^*} \bar{\Psi}_2(\tau_i, x, y, \theta, x_k, y_j, \theta_l, x^*_k, y^*_j, \theta^*_l)
= \calG(\tau_i, x, y, \theta) - \frac{\bar{\mu}^2_x(\tau_i, x,y,\theta)}{2 \gamma \sigma^2}. \nonumber
\end{align}
Here, $c_{i,kjl}$ are the RBF coefficients to be determined, $\{ x_k, y_j, \theta_l: \ k=1,\ldots,N_{i,k}, \ j=1,\ldots,N_{i,j}, \ l=1,\ldots,N_{i,l} \}$ is a set of collocation points in the $x, y, \theta$ directions, $\bar{\Psi}_1(\tau_i, x, x_k, y, y_j, \theta, \theta_l)$ represents the part of $\bar{\Psi}$ which is linear in $\calC$ and $\bar{\Psi}_2(\tau_i, x, y, \theta, x_k, y_j, \theta_l, x^*_k, y^*_j, \theta^*_l)$ is a similar function with quadratic terms from $\bar{\Psi}$. The primed indices $\{ k^*, j^*, l^* \}$ in the double sum follow the same ranges as $\{ k, j, l \}$ and are used to mark the RBF coefficients in the quadratic terms in \eqref{VolterraRBF}
\begin{align} \label{defPsi12}
\bar{\Psi}_1&(\tau_i, x, y, \theta, x_k, y_j, \theta_l) = \bar{\eta} \calC_x(\bm{q}, \bm{c}) + \mu_y(\tau_i, \bm{q}) \calC_y(\bm{q}, \bm{c}) + \mu_\theta(\theta) \calC_\theta(\bm{q}, \bm{c}) - r \calC(\bm{q}, \bm{c}), \\
\bar{\Psi}_2&(\tau_i, x, y, \theta, x_k, y_j, \theta_l, x^*_k, y^*_j, \theta^*_l) = \frac{1}{2} \gamma \sigma_y^2 \calC^2_y(\bm{q},\bm{c}, \bm{c}^*) + \frac{1}{2} \gamma \sigma_\theta^2 \calC^2_\theta(\bm{q}, \bm{c}, \bm{c}^*). \nonumber
\end{align}
\vspace{-2em}
\begin{align*}
\calC(\bm{q}, \bm{c}) &= e^{-\varepsilon \left[ (x-x_k)^2 + (y-y_j)^2 + (\theta - \theta_l)^2\right]},  \qquad \bm{q} = \{ x,y,\theta\},
\qquad \bm{c} = \{ x_k,y_j,\theta_l \}, \\
\calC^2(\bm{q},\bm{c}, \bm{c}^*) &= e^{-\varepsilon \left[ (x-x_k)^2 + (y-y_j)^2 + (\theta - \theta_l)^2\right]} e^{-\varepsilon \left[ (x-x_{k^*})^2 + (y-y_{j^*})^2 + (\theta - \theta_{l^*})^2\right]}. \nonumber
\end{align*}
Computation of $\calK(\tau_m, x, y, \theta; \calC)$ is detailed in \cref{appMatrixRBF}.

When $x, y, \theta$ are chosen to coincide with the collocation points, \eqref{VolterraRBF} yields a system of nonlinear (quadratic) equations for $c_{i,k,j,l}$, which can be expressed in a matrix form
\begin{align} \label{matrixRBF}
\bm{A} \cdot \bm{c} &= \bm{G} + \bm{B} \cdot \bm{c}^T \otimes \bm{c}.
\end{align}
Here, $\bm{c}$ is a column vector containing the RBF coefficients, $\bm{A}$ is a matrix formed by evaluating  the first sum in \eqref{VolterraRBF} at all combinations of the collocation points $(x_k, y_j, \theta_l)$, $\bm{B}$ is a matrix formed similarly by evaluating the double sum in \eqref{VolterraRBF}, and $\bm{G}$ is a column vector obtained by evaluating $\calG(\tau_i,x,y,\theta)$ at the collocation points $x_k,y_j,\theta_l$. The explicit representation of all these matrices can be found in \cref{appMatrixRBF}.

This nonlinear system can be solved using iterative methods such as:
\begin{itemize}
\item  Newton-Raphson method, which has a computational complexity of $O((N_k N_j N_L)^3)$ per iteration;
\item Fixed point iteration method, that can be expressed, e.g., as follows
\begin{align}
\bm{c}_{n+1} &= \bm{A}^{-1} [\bm{G} + \bm{B} \cdot \bm{c}^T_{n} \otimes \bm{c}_{n}],
\end{align}
\noindent where $c_n$ represents the solution vector at the $n$-th iteration. The iterations begin with an initial guess $\bm{c}_0$ and continues until reaching a specified convergence tolerance.
\end{itemize}
Since such a system must be solved at each time step $\tau_i$ for $i \in [1,N]$, the total computational complexity is $O((N_k N_j N_L)^3 N M)$, where $M$ denotes the average number of iterations required for convergence.

\section{An efficient approach using operator splitting as an alternative method} \label{sec-solSplit}

The method developed in \cref{solRBF} for solving the Volterra integral equation relies on a numerical approximation of the temporal integral using quadrature rules. By leveraging the RBF method with Gaussian kernels, all integrands arising from this temporal discretization can be evaluated in closed form, which is a significant advantage of this approach. However, at each time step, a nonlinear (quadratic) system of equations must be solved iteratively. While the number of iterations typically ranges between 10 and 20, it can exceed this range in some cases, leading to slower computational performance.

An alternative approach, similar in spirit but distinct in implementation, can be developed using the operator splitting technique. This method replaces the need to solve a nonlinear system with the solution of three linear systems of equations. From a computational standpoint, this is equivalent to performing only three iterations in the method of \cref{solRBF}, potentially offering a significant improvement in efficiency.

The core idea of the operator splitting method is as follows. We solve the problem in \eqref{SemiLinearPDE} using the same temporal grid
$\Gamma_i$ as in \cref{solRBF}. At each time step $\tau_i$  the model coefficients $c(t), a_1(t), a_2(t)$ in \eqref{Marketron_3D,fhparam} are assumed to be constant, typically set to their mean value at the points $\tau_i$ and $\tau_{i-1}$. Consequently, the problem in \eqref{SemiLinearPDE} is solved at time  $\tau_i$ based on its solution at time $\tau_{i-1}$ (so, over the interval $(\tau_{i-1}, \tau_i = \tau_{i-1} + \Delta \tau_i]$\, ) , where all model coefficients are treated as time-independent.

For a general approach to splitting techniques for {\it linear} operators using Lie algebras, we refer the reader to \cite{LanserVerwer,ItkinBook}. Let us represent \eqref{SemiLinearPDE} in the form
\begin{align} \label{splitEq}
\fp{\calC(\tau, \bm{p}}{\tau} &= \calL(\calC, \bm{p}), \\
\calL(\calC, \bm{p}) &= \frac{1}{2}\sigma^2 \sop{\calC}{x} + \frac{1}{2}\sigma_y^2 \sop{\calC}{y} + \frac{1}{2}\sigma_\theta^2 \sop{\calC}{\theta} + \Phi, \nonumber
\end{align}
\noindent where $\calL(\calC, \bm{p})$ is the differential operator from \eqref{SemiLinearPDE}.

Decomposing the total (compound)  operator $\calL$ is natural when $\calL$ can be represented as a sum of $k$ operators $\sum_{m} \calL_m$. Since these operators do not explicitly depend on time, the formal solution of \eqref{splitEq} would be
\begin{align} \label{solSplit}
\calC(\tau_i, \bm{p}) &= e^{ \Delta \tau_i (\calL_1 + \calL_2 + \calL_3)} \calC(\tau_{i-1},\bm{p}), \\
\calL_1(\tau_i, \bm{p}) &= \frac{1}{2}\sigma^2 \sop{\calC}{x} + \bar{\eta} \fp{\calC}{x} -r \calC - \frac{\bar{\mu}^2_x}{2 \gamma \sigma^2}, \nonumber \\
\calL_2(\tau_i, \bm{p}) &= \frac{1}{2}\sigma_y^2 \sop{\calC}{y} + \mu_y \fp{\calC}{y} + \frac{1}{2} \gamma \sigma_y^2 \left( \fp{\calC}{y} \right)^2, \nonumber \\
\calL_3(\tau_i, \bm{p}) &= \frac{1}{2}\sigma_\theta^2 \sop{\calC}{\theta} + \mu_\theta \fp{\calC}{\theta} + \frac{1}{2} \gamma \sigma_\theta^2 \left( \fp{\calC}{\theta} \right)^2. \nonumber
\end{align}

If all operators $\calL_m$ commute, this expression can be  factorized into a product of operators
\begin{equation} \label{factor}
e^{ \Delta \tau_i (\calL_1 + \calL_2 + \calL_3)} = e^{ \Delta \tau_i \calL_1} e^{ \Delta \tau_i \calL_2} e^{ \Delta \tau_i \calL_3}.
\end{equation}
Then, \eqref{splitEq} can be solved in three steps sequentially by the following procedure:
\begin{align} \label{solSteps}
\calC^{(1)}(\tau_i, \bm{p}) &= e^{\Delta \tau_i \calL_1} \calC(\tau_{i-1}, \bm{p}), \\
\calC^{(2)}(\tau_i, \bm{p}) &= e^{\Delta \tau_i \calL_2} \calC^{(1)}(\tau_i, \bm{p}), \nonumber \\
\calC(\tau_i, \bm{p}) &= e^{\Delta \tau_i \calL_3} \calC^{(2)}(\tau_i, \bm{p}). \nonumber
\end{align}
This algorithm is exact (no bias) if all the operators commute.

However, since in \eqref{solSplit} the coefficients of operators $\calL_i$ are functions of the independent variables $x, y, \theta$, these operators do not commute. In this case a simple splitting in \eqref{solSteps} provides only a first-order approximation in time (i.e., $O(\Delta \tau_i)$) to the exact solution. Higher-order accuracy can be achieved using various approximations based on the Baker-Campbell-Hausdorff formula \cite{BCH}. For example, Strang's splitting provides a second-order approximation in $\Delta \tau_i$ (see \cite{LanserVerwer,ItkinBook} and references therein)\footnote{Since the option payoff depends solely on $x$, it is natural to employ operator $\calL_1$ as the outer layer in this splitting scheme.}
\begin{align} \label{strang1}
e^{ \Delta \tau_i (\calL_1 + \calL_2 + \calL_3)} = e^{ \frac{1}{2}\Delta \tau_i \calL_1} e^{ \frac{1}{2} \Delta \tau_2 \calL_2} e^{ \Delta \tau_i \calL_3} e^{ \frac{1}{2} \Delta \tau_i \calL_2} e^{ \frac{1}{2} \Delta \tau_i \calL_1} + O((\Delta \tau_i)^2).
\end{align}
For parabolic equations with coefficients independent of $\tau$, this composite algorithm achieves second-order accuracy in $\dtau_i$, provided that the numerical procedure solving the corresponding PDE at each splitting step maintains at least second-order accuracy in time.
It can be easily validated that this condition is satisfied in our case, as we employ trapezoidal quadrature for time integration, and also approximate the time-dependent coefficient $a_1(\tau)$ in \eqref{fhparam} at the interval $[\tau_{i-1}, \tau_i]$ by its mean value (and same for $a_2(\tau$)).

The factorized solution of \eqref{splitEq} given by equations \eqref{solSteps} with allowance for \eqref{solSplit} can be transformed back
into a system of PDEs as follows
\begin{alignat}{2} \label{systemPDE}
\fp{\calC^{(1)}(t, \bm{p})}{t} &+ \calL_1(\calC^{(1)}(t, \bm{p})) = 0, &\qquad \calC^{(1)}(t_{i-1}, \bm{p}) &= \calC(t_{i-1}, \bm{p}), \\
\fp{\calC^{(2)}(t, \bm{p})}{t} &+ \calL_2(\calC^{(2)}(t, \bm{p})) = 0, &\qquad \calC^{(2)}(t_{i-1}, \bm{p}) &= \calC^{(1)}(t_{i-1}, \bm{p}), \nonumber \\
\fp{\calC(t, \bm{p})}{t} &+ \calL_3(\calC(t, \bm{p})) = 0, &\qquad \calC(t_{i-1}, \bm{p}) &= \calC^{(2)}(t_{i-1}, \bm{p}). \nonumber
\end{alignat}

The first equation in \eqref{systemPDE} is linear in $\calC$ and can be solved by the method described in \cref{solRBF}. To recall, the method first transforms the PDE into a Volterra integral equation and then solves it using an RBF method with Gaussian kernels. Since each equation in \eqref{systemPDE} is one-dimensional with the other two independent variables fixed (dummy), the corresponding Green's function takes a modified form
\begin{align} \label{GreenSplit}
G(\tau; x,y,\theta|\xi, \eta, \zeta) = \frac{1}{(2 \pi \tau)^{1/2} \sigma} e^{- \frac{(x-\xi)^2}{2 \sigma^2 \tau} } \delta(y - \eta) \delta(\theta - \zeta).
\end{align}

The unknown RBF coefficients can be obtained by solving a system of {\it linear} equations (here we should set $\bar{\Psi}_2 = 0$), where all matrices can be computed in closed form as shown in \cref{appMatrixRBF}. Observe that this system still contains all 3D coefficients $c_{i,j,k}$ despite the variables $y$ and $\theta$ being dummy variables in this context.

So far, everything looks pretty similar to the method of \cref{solRBF}. However, things change (and this is the core idea of using the operator splitting method) when we proceed to solving the second and the third PDEs in \eqref{systemPDE}. For instance, consider the second PDE in \eqref{systemPDE}. It is nonlinear, having a quadratic nonlinearity in the first spatial derivative $\calC_y$. However, since in this context the variables $x$ and $\theta$ are dummy, a standard Cole-Hopf transformation, \cite{GrasselliHurd2007,HH2008,HalperinItkin2014} can be applied to reduce this equation to a {\it linear} PDE, namely
\begin{equation}
\fp{w}{t} + \frac{1}{2}\sigma_y^2 \sop{w}{y} + \mu_y \fp{w}{y} = 0, \qquad \calC(t,\bm{p}) = \frac{1}{\gamma} \log [C_N w(t, \bm{p})],
\end{equation}
\noindent where $C_N > 0$ is some normalization constant. This linear PDE can be solved in the same way as the first PDE in \eqref{systemPDE}, i.e., by using the approach of \cref{solRBF}. In a similar way, the last PDE in \eqref{systemPDE} can be solved as well.

This, however, requires two extra steps to convert the RBF coefficients $\bm{c}$ obtained at the first step into those which are used in the initial condition for the second step. If the same Gaussian kernel is used for solving the transformed PDE for the dependent variable $w(t, \bm{p})$, then the new RBF coefficients $\bm{c}_w$ solve the following linear system of equations
\begin{align} \label{matrixRBFtr}
\bm{\Theta} \cdot \bm{c}_w &= \frac{1}{C_N} \exp \left( \gamma \bm{\Theta} \cdot \bm{c} \right).
\end{align}
Here, matrix $\bm{\Theta}$ is defined in \eqref{gauss} and is constructed by using a set of collocation points $\{x_k, y_j, \theta_l\}$.

Accordingly, after the last step of splitting is done, the inverse transformation is applied to obtain the RBF coefficients $\bm{c}$ as the solution of the linear system
\begin{align} \label{matrixRBFtrI}
\bm{\Theta} \cdot \bm{c} &= \frac{1}{\gamma} \log \left( C_N (\bm{\Theta} \cdot \bm{c}_w)^+ \right).
\end{align}
Since all elements of the vector $\bm{\Theta} \cdot \bm{c}$ represent Call option prices corresponding to various sets of $\{x_k, y_j, \theta_l\}$, they should be nonnegative. This is achieved by using $(\cdot)^+$ in \eqref{matrixRBFtr}.

The idea behind introducing the normalization constant $C_N$ is as follows. In \eqref{matrixRBFtr}, if $\gamma$ and the option prices are high, exponentiation gives rise to very large values which could exceed the maximum floating point representation supported by some particular programming language or computer architecture. In particular, in Python, since the \verb|scipy.sparse.linalg.minres| function uses float32 (as it is taken from LAPACK written in C), this maximum exponent is approximately $E_{\max} = 350$. Accordingly, since the SPX prices could be of the order of 200, this imposes serious restrictions on the values of $\gamma$. However, we can address this by taking
\begin{align}
C_N &= e^{A}, \qquad A =
\begin{cases}
0, & \max[ \gamma \bm{\Theta} \cdot \bm{c}] < E_{\max}, \\
\max[\gamma \bm{\Theta} \cdot \bm{c}] - E_{\max}, & \max[ \gamma \bm{\Theta} \cdot \bm{c}] > E_{\max}.
\end{cases}
\end{align}

This transforms our scheme in \cref{matrixRBFtr,matrixRBFtrI} to:
\begin{align} \label{matrixRBFtrNorm}
\bm{\Theta} \cdot \bm{c}_w &= \exp \left( \gamma \bm{\Theta} \cdot \bm{c} - A\right), \qquad \bm{\Theta} \cdot \bm{c} = \frac{1}{\gamma} \left[ \log (\bm{\Theta} \cdot \bm{c}_w)^+ + A \right],
\end{align}
\noindent which eliminates the issue with handling large numbers.

It is important to note that in our splitting scheme, we allocated the source term to the first step of splitting. This allowed us to apply the Cole-Hopf transformation at the second and third steps since the corresponding equations don't contain the source term. However, the source term can also be distributed among all operators $\calL_i$, which typically improves numerical accuracy. This improvement occurs because while the source term depends on all independent variables, each splitting step operates on only one variable, treating the others as fixed (dummy) parameters. If this approach is taken, the Cole-Hopf transformation can still be applied, but in this case, the modified PDE will also contain a nonlinear term. Nevertheless, this PDE can be solved by the RBF method in the same way as discussed above.

The system of PDEs in \eqref{systemPDE} can also be solved using FD methods. The key advantage of the RBF method is that it requires significantly fewer collocation points in the $y$ and $\theta$ dimensions — approximately 5 points suffice, whereas FD methods typically need an order of magnitude more points for comparable accuracy. Moreover, the only part of \eqref{Volterra} which depends on the strike $K$ is $\mathcal{I}(\tau,x)$. Therefore, linear systems of equations that appear in our RBF method having various right-hand sides, can be solved simultaneously. In other words, obtaining the solution requires solving just one linear system with multiple right-hand sides, which can be efficiently done with modern software. Thus, pricing options with different strikes but the same other parameters can be obtained in one sweep. Consequently, our method offers substantial computational efficiency, with execution times about two orders of magnitude faster than traditional FD approaches. Moreover, solving forward Kolmogorov PDEs rather than the backward PDEs shown in equation \eqref{systemPDE} would provide enhanced computational efficiency as options with various maturities could be priced simultaneously, \cite{ItkinBook}. However, this approach lies beyond the scope of the present paper.

\subsection{Numerical experiments} \label{exper}

In this numerical experiment, we price Call options on an underlying asset whose time evolution follows the Marketron model in \eqref{Marketron_3D}. The parameters of these options (which mimic those of SPY) are given in Table~\ref{callParam}. We use all combinations of $S$ and $K$, resulting in a total of 49 options to price.

To solve the Volterra equation in \eqref{trapezoid} by using the RBF method with splitting as described in \cref{sec-solSplit}, we choose a uniform grid in $t \in [0,T]$ with $N_\tau = 30$, and set the collocation points in $x, y, \theta$ directions using $N_x = 20, N_y = 5, N_\theta = 5$ uniformly distributed over the intervals
\begin{align}
x_k &\in [\min(0.7 x_{\min}, -0.5),  \max(1.3 x_{\max}, 0.5)], \quad k \in [1,N_x], \\
x_{\min} &= \min[\log(S_i/S_*)], \, x_{\max} = \max[\log(S_i/S_*)], \, i=1,\ldots,7, \nonumber \\
y_j &\in [\min(0.7 y_{\min}, -0.5),  \max(1.3 y_{\max}, 0.5)], \quad j \in [1,N_y], \nonumber \\
\theta_l &\in [\min(0.7 \theta_{\min}, -0.5),  \max(1.3 \theta_{\max}, 0.5)], \quad l \in [1,N_\theta]. \nonumber
\end{align}
It is important to recognize that in this setting the initial values at time $t=0$ of $y$ and $\theta$ should be added to the calibration parameters of the model since variables $y_t, \theta_t$ are unobservable. A close analogy could be the initial instantaneous variance $v_0$ in the Heston model which also is found by calibration, \cite{Balaraman2016}. Here, we take them as $y = 0.1, \theta = 0.5$.

\begin{table}[!htb]
\centering
\begin{tabular}{|l|r|r|r|}
\toprule
\rowcolor[rgb]{ .949,  .808,  .937} \multicolumn{1}{|c|}{S, K} & \multicolumn{1}{c|}{T} & \multicolumn{1}{c|}{r} & \multicolumn{1}{c|}{q} \\
\midrule
950, 975, 985, 1000, 1015, 1025, 1050 & 0.25  & 0.01  & 0.005 \\
\bottomrule
\end{tabular}%
\caption{Parameters of Call options used in the test.}
\label{callParam}%
\end{table}%

Following findings in \cite{CarrItkinMuraveyHeston}, we implement a \verb+minres+ iterative solver to handle the system of linear equations obtained via the RBF method. This solver is particularly effective for matrices that are symmetric but not positive definite. Although our non-standard (non-Gaussian) RBF technically produces a non-symmetric matrix, our experiments show it is nearly symmetric, with maximum absolute differences between corresponding elements $|a(i,j) - a(j,i)| \approx 1.e-4$. After testing various iterative solvers, we found that \verb+minres+ consistently delivered superior results. Its key advantage is the ability to construct an orthogonal basis for the Krylov subspace using three-term recurrence relations \cite{minres1975}.

We observe that due to rounding errors, some eigenvalues of our RBF matrices are either very small negative numbers or zero. Ideally, the RBF matrix should undergo standard regularization procedures (which we also utilize to convert it, if necessary, to a nearest symmetric positive semi-definite matrix), \cite{Higham1998}. However, since this error is related to the quality of the RBF interpolation, more stable methods than the global RBF approach would likely yield better accuracy. Indeed, for the Gaussian RBF method, a typical matrix in \eqref{gauss} has a condition number of approximately $10^{19}$, causing even iterative solving methods to produce larger errors than would occur with a well-conditioned matrix.

To complete the marketron model setting, we use the model parameters from Table~1 of \cite{HalperinItkin2025ar}, but since the instantaneous interest rate $r$ here is fixed, the value of the parameter $\bar{\eta} = r - \sigma^2/2$ is replaced accordingly. Additionally, we need to add the values of $\gamma$ and $\epsilon$ to the model parameters. Overall, all these values are presented in Table~\ref{calibParam}.
\begin{table}[tbhp]
\begin{center}
\scalebox{0.87}{
\begin{tabular}{|l|r|r|r|r|r|r|r|r|r|r|r}
\toprule
\rowcolor[rgb]{ .792,  .929,  .984} {\bf parameter} & $\bm \sigma$ & $\bm \sigma_y$ & $\bm \sigma_z$ & $\bm \gamma$ & $\bm k$ & $\bm \mu$ & $\bm g$ & ${\bm \hat{\theta}}$ & ${\bm \bar{y}}$ & ${\bm y}$ \\ \hline
value       & 0.37 & 0.3800 & 0.8334 & 0.2 & 1.2869 & 1.6671 & 0.6831 & 6.7865 & 0.4731 & 0.1  \\ \hline
\rowcolor[rgb]{ .792,  .929,  .984} {\bf parameter} & $\bm c$ & $\bm b_1$ & $\bm b_2$ & $\bm k_{1,x}$ & $\bm k_{2,x}$ & $\bm k_{3,x}$ & $\bm k_{1,y}$ & $\bm k_{2,y}$ & $\bm k_{3,y}$ & ${\bm \theta}$ \\ \hline
value       & 3.9305 & 1.6819 & -1.2102 & -3.2002 & 2.7417 & -1.8832 & -0.7855 & 3.8901 & 1.5588 & 0.5 \\ \hline
\bottomrule
\end{tabular}
}
\caption{Parameters of the model in \eqref{Marketron_3D} found in \cite{HalperinItkin2025ar} by calibration to S\&P500 weekly returns from 2000 to Sept. 2024.}
\label{calibParam}
\end{center}
\end{table}

\begin{figure}[!htb]
\vspace{-1.4em}
\begin{center}
\includegraphics[width=0.6\textwidth]{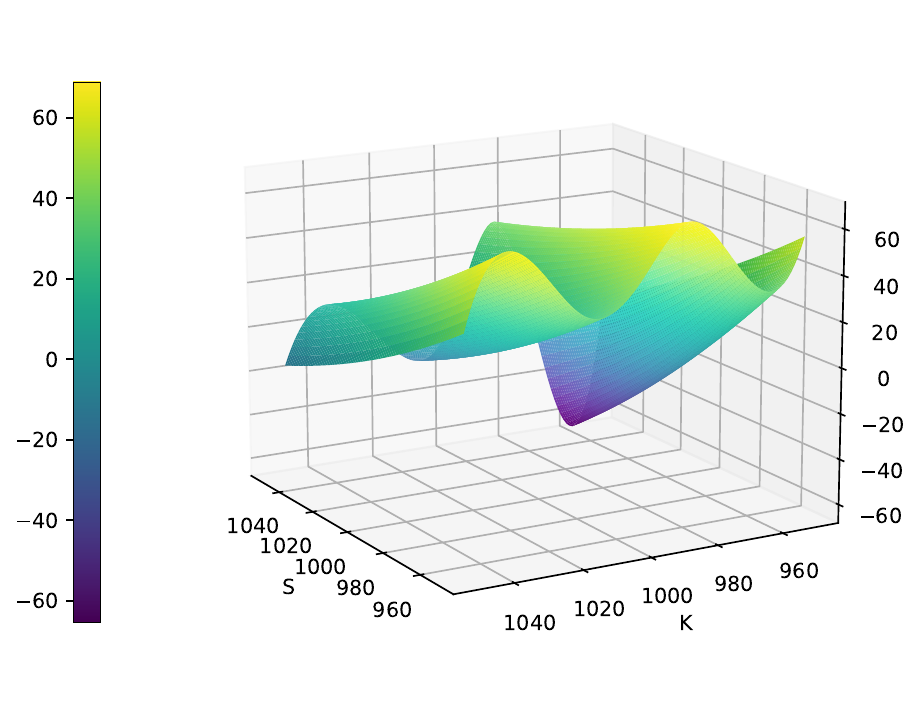}
\end{center}
\vspace{-2em}
\caption{The difference between the Call option prices in Table~\ref{callRes} and those computed by using the Black-Scholes formula.}
\label{callDiff}
\end{figure}

The results of this test, where we set $\epsilon = 0.2$ and adjust the volatility to $\sigma = 0.37$, are presented in Table~\ref{callRes}. Meanwhile, Fig.~\ref{callDiff} illustrates the difference in Call prices between the Marketron model with these parameters and those computed using the Black-Scholes formula. Notably, the Marketron option prices exhibit highly nonlinear behavior compared to the Black-Scholes prices.

Furthermore, our experiments demonstrate that the option price is highly sensitive to the values of $\sigma$ and $\epsilon$. To highlight this, Table~\ref{callRes2} displays Call option prices where $\epsilon$ remains unchanged, but $\sigma$ is reverted to its original value from Table~\ref{callParam}. As shown, the magnitude of the option prices increases significantly under this adjustment. However, increasing $\epsilon$ partially mitigates this effect. Consequently, typical values for these parameters should be determined by calibrating the option prices to market data.

\begin{table}[htbp]
\centering
\scalebox{0.9}{
\begin{tabular}{|c|r|r|r|r|r|r|r|}
\toprule
\multicolumn{8}{|c|}{\cellcolor[rgb]{ .792,  .929,  .984}\textbf{K}} \\
\midrule
\rowcolor[rgb]{ .792,  .929,  .984} \textbf{S } & \textbf{950.00} & \textbf{975.00} & \textbf{985.00} & \textbf{1000.00} & \textbf{1015.00} & \textbf{1025.00} & \textbf{1050.00} \\
\hline
\rowcolor[rgb]{ .855,  .949,  .816} \textbf{950.00} & 126.33 & 116.61 & 124.29 & 94.77 & 79.59 & 92.10 & 69.94 \\
\hline
\textbf{975.00} & 122.18 & 111.63 & 125.96 & 89.01 & 73.40 & 85.43 & 62.46 \\
\hline
\rowcolor[rgb]{ .855,  .949,  .816} \textbf{985.00} & 121.55 & 110.68 & 127.51 & 87.77 & 71.99 & 83.81 & 60.53 \\
\hline
\textbf{1000.00} & 121.66 & 110.30 & 130.71 & 86.97 & 70.96 & 82.45 & 58.74 \\
\hline
\rowcolor[rgb]{ .855,  .949,  .816} \textbf{1015.00} & 122.98 & 111.14 & 134.94 & 87.41 & 71.20 & 82.32 & 58.21 \\
\hline
\textbf{1025.00} & 124.51 & 112.34 & 138.31 & 88.37 & 72.02 & 82.89 & 58.52 \\
\hline
\rowcolor[rgb]{ .855,  .949,  .816} \textbf{1050.00} & 130.48 & 117.52 & 148.55 & 92.96 & 76.31 & 86.52 & 61.55 \\
\bottomrule
\end{tabular}%
}
\caption{Call option prices computed by using the Marketron model with parameters in Table~\ref{callParam} (except $\sigma = 0.37$) and $\epsilon = 0.2$.}
\label{callRes}%
\end{table}%
\begin{table}[!htb]
\centering
\scalebox{0.9}{
\begin{tabular}{|c|r|r|r|r|r|r|r|}
\toprule
\rowcolor[rgb]{ .792,  .929,  .984} \multicolumn{8}{|c|}{\textbf{K}} \\
\midrule
\rowcolor[rgb]{ .792,  .929,  .984} \textbf{S } & \textbf{950.00} & \textbf{975.00} & \textbf{985.00} & \textbf{1000.00} & \textbf{1015.00} & \textbf{1025.00} & \textbf{1050.00} \\
\hline
\rowcolor[rgb]{ .855,  .949,  .816} \textbf{950.00} & \cellcolor[rgb]{ 1,  1,  1}322.24 & \cellcolor[rgb]{ 1,  1,  1}310.75 & \cellcolor[rgb]{ 1,  1,  1}306.26 & \cellcolor[rgb]{ 1,  1,  1}299.48 & \cellcolor[rgb]{ 1,  1,  1}292.61 & \cellcolor[rgb]{ 1,  1,  1}288.01 & \cellcolor[rgb]{ 1,  1,  1}276.75 \\
\hline
\textbf{975.00} & \cellcolor[rgb]{ .855,  .949,  .816}335.50 & \cellcolor[rgb]{ .855,  .949,  .816}323.42 & \cellcolor[rgb]{ .855,  .949,  .816}318.69 & \cellcolor[rgb]{ .855,  .949,  .816}311.57 & \cellcolor[rgb]{ .855,  .949,  .816}304.35 & \cellcolor[rgb]{ .855,  .949,  .816}299.51 & \cellcolor[rgb]{ .855,  .949,  .816}287.67 \\
\hline
\rowcolor[rgb]{ .855,  .949,  .816} \textbf{985.00} & \cellcolor[rgb]{ 1,  1,  1}340.76 & \cellcolor[rgb]{ 1,  1,  1}328.46 & \cellcolor[rgb]{ 1,  1,  1}323.64 & \cellcolor[rgb]{ 1,  1,  1}316.38 & \cellcolor[rgb]{ 1,  1,  1}309.02 & \cellcolor[rgb]{ 1,  1,  1}304.09 & \cellcolor[rgb]{ 1,  1,  1}292.03 \\
\hline
\textbf{1000.00} & \cellcolor[rgb]{ .855,  .949,  .816}348.63 & \cellcolor[rgb]{ .855,  .949,  .816}335.99 & \cellcolor[rgb]{ .855,  .949,  .816}331.03 & \cellcolor[rgb]{ .855,  .949,  .816}323.57 & \cellcolor[rgb]{ .855,  .949,  .816}316.01 & \cellcolor[rgb]{ .855,  .949,  .816}310.95 & \cellcolor[rgb]{ .855,  .949,  .816}298.54 \\
\hline
\rowcolor[rgb]{ .855,  .949,  .816} \textbf{1015.00} & \cellcolor[rgb]{ 1,  1,  1}356.45 & \cellcolor[rgb]{ 1,  1,  1}343.48 & \cellcolor[rgb]{ 1,  1,  1}338.39 & \cellcolor[rgb]{ 1,  1,  1}330.72 & \cellcolor[rgb]{ 1,  1,  1}322.97 & \cellcolor[rgb]{ 1,  1,  1}317.77 & \cellcolor[rgb]{ 1,  1,  1}305.04 \\
\hline
\textbf{1025.00} & \cellcolor[rgb]{ .855,  .949,  .816}361.64 & \cellcolor[rgb]{ .855,  .949,  .816}348.45 & \cellcolor[rgb]{ .855,  .949,  .816}343.27 & \cellcolor[rgb]{ .855,  .949,  .816}335.48 & \cellcolor[rgb]{ .855,  .949,  .816}327.59 & \cellcolor[rgb]{ .855,  .949,  .816}322.31 & \cellcolor[rgb]{ .855,  .949,  .816}309.35 \\
\hline
\rowcolor[rgb]{ .855,  .949,  .816} \textbf{1050.00} & \cellcolor[rgb]{ 1,  1,  1}374.52 & \cellcolor[rgb]{ 1,  1,  1}360.80 & \cellcolor[rgb]{ 1,  1,  1}355.41 & \cellcolor[rgb]{ 1,  1,  1}347.29 & \cellcolor[rgb]{ 1,  1,  1}339.08 & \cellcolor[rgb]{ 1,  1,  1}333.59 & \cellcolor[rgb]{ 1,  1,  1}320.10 \\
\bottomrule
\end{tabular}%
}
\caption{Call option prices computed by using the Marketron model with parameters in Table~\ref{callParam} and $\epsilon = 0.2$.}
\label{callRes2}%
\end{table}%

All these results are obtained in Python 3.11 using PC with two Intel Quad-Core i7-4790 CPUs, each running at 3.80 GHz. A typical elapsed time to compute 49 prices presented in Table~\ref{callRes} is 0.7 seconds. This efficiency is due to two factors: i) option prices with the same maturity $T$ but different strikes $K$ are computed simultaneously since at every time step we solve a single linear system of equations with multiple right-hand sides, and ii) once solved, we obtain an RBF interpolator, so various stock/index prices can be substituted into it to obtain all option prices corresponding to these values of $S$ at once.

\section{Calibration of the Marketron model to market option data} \label{sect:calib}

While \cref{sect:indiff} demonstrates how to price options in the Marketron model under incomplete market conditions, this approach remains computationally intensive for direct use in calibration. Although due to natural internal parallelism, all option prices can be computed in parallel at each calibration step, the authors lack access to the necessary computational resources to leverage this capability effectively.

In this section, we present a practical implementation of our calibration
method based on global optimization with suitable constraints.
Our method is this section relies on a minor change in representing signals in the Marketron model, which is introduced to improve computational efficiency. We will first present this reformulation, and then provide details of our global optimization method.


\subsection{A reduced approach to modelling signals $\theta$} \label{redApp}

In \cite{HalperinItkin2025ar}, it was revealed that if in \eqref{Marketron_3D} the condition $f(\theta_t) + \bar{\eta} < 0$ holds, the log-price $x_t$ tends to negative infinity as $t \rightarrow \infty$, whereas in the opposite case, it asymptotically approaches positive infinity. Both scenarios are expected to have very low probability within the time horizons considered in that paper, and this also holds true for options. While such behavior would be anticipated for the "true" set of model parameters in a realistic setting, these parameters are initially unknown and must be determined through calibration. This poses a potential challenge for the calibration method proposed in \cite{HalperinItkin2025ar}, which relies on simulated trajectories and non-linear filtering. With arbitrary parameter values, an excessive number of trajectories may diverge toward negative or positive infinity. To mitigate this issue, we introduced a more general specification of the signals using inhomogeneous functions $f$ and $h$, which can change sign over time. These functions are explicitly defined in \eqref{fhparam}.

Although this specification aligns with our objectives, it introduces eight additional calibration parameters, increasing computational complexity. However, our primary goal was to achieve signals with bounded variation. This objective, however, can be accomplished in many different ways, for instance, by employing the concept of bounded diffusion among others. Therefore, in this section our main goal is  to propose an alternative tractable model for the marketron signals which is of a lower computational complexity

In the financial literature, bounded diffusion is widely used for modeling various processes, such as stochastic correlation and other bounded dynamics, as discussed in \cite{Zetocha2015, Itkin2017, Domingo2020} and references therein. A common approach involves applying a transformation $\rho_t = g(X_t)$, where $X_t$ is an arbitrary diffusion process (possibly driftless), and the function $g(X_t)$ maps the support of $X_t$ to the interval $[-1, 1]$, \cite{Emmerich2006}. Popular choices for such transformations include the hyperbolic tangent \cite{Teng2016}, the normal cumulative distribution function (CDF) \cite{Carr2017}, and the normalized inverse tangent \cite{Emmerich2006}, among others. In principle, any continuous mapping from $\mathbb{R}$ to $[-1, 1]$ can be employed, with the selection often guided by additional desirable properties such as tractability. While this approach may be less intuitive, it enables the development of sophisticated models for stochastic correlations. Notably, our model in \eqref{fhparam} falls directly within this class of transformations.

Alternatively, one can employ stochastic processes that are inherently bounded by definition, such as the Jacobi process, which has been used in \cite{Emmerich2006, Zetocha2015}. Adopting this approach, we can define our stochastic signal $\theta_t$, e.g., as:
\begin{equation} \label{jac}
d \theta_t = k(\hat{\theta} - \theta_t) dt + \sigma_\theta \sqrt{(\theta_{\max} - \theta_t)(\theta_t - \theta_{\min})} dZ_t.
\end{equation}
Here, $ \theta_{\min} $ and $ \theta_{\max} $ represent the lower and upper bounds of the process, which are assumed to be known. For simplicity, we set $ \theta_{\min} = -\theta_{\max} $ in this paper. Since the process $ \theta_t $ in \eqref{jac} is inherently of bounded variation, we can redefine the functions $ f(\theta_t) $ and $ h(\theta_t) $ by dropping additional transformations and see them as linear functions of the signal
\begin{equation}
f(\theta_t) = b_1 \theta_t, \qquad h(\theta_t) = b_2 \theta_t,
\end{equation}
\noindent where $ b_1 $ and $ b_2 $ are constants. This simplification reduces the number of the corresponding Marketron parameters from eight to three ($b_1, b_2, \theta_{\max}$), thus significantly enhancing computational efficiency. From a financial standpoint, this change implies that when the signal $\theta_t$ reaches its mean-reversion level, its contribution to the spot price and the memory variable $y$ remains constant thereafter. In contrast, under the previous parameterization in \eqref{fhparam}, this contribution was time-dependent.

However, since the diffusion is already bounded, its domain of definition is a finite interval $\theta_t \in (\theta_{\min}, \theta_{\max})$. In such a case, solving an HJB equation by the method proposed in \cref{sec:Volterra} becomes more challenging because the Green's function for the heat equation defined on a finite interval has a more complex form (various examples can be found in \cite{ItkinLiptonMuraveyBook}). Consequently, computing integrals similar to those in \cref{appB} becomes more difficult. Therefore, we continue with the transformation method as before, but simplify the transformation function by setting
\begin{equation} \label{defTanh}
f(\theta_t) = b_1 \cos(\theta_t), \qquad h(\theta_t) = b_2 \sin(\theta_t).
\end{equation}
These definitions preserve the $\theta$-dependent part of the Green's function in the same form while ensuring that the other integrals (convolutions of the Green's function with the remaining source terms) remain tractable.

Indeed, if for pricing options we use the splitting method described in \cref{sec-solSplit}\footnote{It can be shown that for the method described in \cref{sec:Volterra} the convolutions can also be computed analytically, but we omit these details here.}, there are two instances where a convolution of $f(\zeta)$ or $h(\zeta)$ with the Green's function of the problem must be computed. The first occurs in the second line of \eqref{systemPDE} where $h(\theta)$ is part of $\mu_y$ defined in \eqref{Marketron_3D}. However, since the Green's function for the second line of \eqref{systemPDE} is
\begin{equation}
G(\tau; x,y,\theta|\xi, \eta, \zeta) = \frac{1}{(2 \pi \tau)^{1/2} \sigma_y} e^{- \frac{(y-\eta)^2}{2 \sigma_y^2 \tau}} \delta(x - \xi) \delta(\theta - \zeta),
\end{equation}
\noindent the corresponding convolution simply reduces to $h(\theta)$. The same logic applies to the convolution of $f(\xi)$ and the Green's function for the first line of \eqref{systemPDE}, which is
\begin{equation}
G(\tau; x,y,\theta|\xi, \eta, \zeta) = \frac{1}{(2 \pi \tau)^{1/2} \sigma} e^{- \frac{(x-\xi)^2}{2 \sigma^2 \tau}} \delta(y - \eta) \delta(\theta - \zeta).
\end{equation}
Hence, this convolution also simply reduces to $f(\theta)$. This tractability represents another advantage of the splitting method proposed in \cref{sec-solSplit}. Thus, with the new parameterization in \eqref{defTanh}, the number of the corresponding Marketron parameters reduces from eight to two ($b_1, b_2$), again significantly enhancing computational efficiency. From the financial standpoint, this change means that when the signal $\theta_t$ reaches its mean-reversion level, its contribution to both the spot price and the memory variable $y$ remains constant, whereas with the previous parameterization in \eqref{fhparam}, it was a function of time.

Additionally, in the subsequent analysis, we aim to examine the underlying stock log-returns generated by simulating our model using parameters calibrated to option data. From this perspective, for time-homogeneous parameters using monotonic functions like $\tanh$ is preferable as compared with non-monotonic functions (e.g., $\cos$, $\sin$) because with our artificial model of signals they introduce less noise into the stock log-returns.

\subsection{Global optimization and parallelization}

In this section, we calibrate the Marketron model to options market data using the direct option prices described in \cref{sec-solSplit}. For a single maturity $T$ and option type (Call or Put), the calibration procedure consists of ten steps - five for computing $ \calC(t, s, y, \theta)$(as in \cref{sec-solSplit}) and five for $\calC^0(t, s, y, \theta)$. At each step, we solve a system of linear equations with multiple right-hand sides corresponding to different strikes. For RBF collocation, we use $N_x = 20$, $N_y = 5$, and $N_\theta = 5$ points, symmetrically and uniformly distributed around the current values of $x$, $y$, and $\theta$, respectively, along with $N_\tau = 30$ time nodes.

Since the modified version of the Marketron model described in this paper assumes constant parameters, we calibrate it independently for each maturity. Although we attempted to calibrate the model to the term structure of option prices (incorporating multiple maturities), the results were insufficiently accurate for further analysis.

For calibration, we use daily OptionMetrics snapshots of Call and Put prices for a given maturity $T$. We restrict strikes $K$ to $S/K < 1.05$ for Puts and $K/S < 1.05$ for Calls, including OTM, ATM, and near-ATM ITM options. All data processing and calculations are performed using the \verb|polars| package, which leverages Rust-based optimizations and lazy evaluation for speed. The pricer operates directly on polars DataFrames via the \verb|map_elements| functionality.

To determine model parameters, we formulate an optimization problem by minimizing the least-squares residuals between computed and market option prices. Initially, we employed the Matlab package \verb|CEopt|, \cite{CunhaJr2024CEopt}, following the approach in \cite{HalperinItkin2025ar} while translated the package to Python, but encountered poor convergence. Therefore, instead, we adopted the \verb|mystic| global optimization package \cite{Mystic2011}, customizing it by:
\begin{itemize}
\item Incorporating our constraint system,
\item Enabling another parallel execution of populations,
\item Implementing a custom monitoring system,
\item Embedding the package within our Python Marketron framework.
\end{itemize}

We use a differential evolution solver with populations per generation set to $10 \times N_p$, where $N_p = 15$ is the number of calibrated parameters. For a typical maturity $T$, calibration involves around 150 options. On a laptop with two Intel Quad-Core i7-4790 CPUs (3.80 GHz), each generation completes in approximately 2 minutes.The observed relative accuracy of calibration is about 8\% if the set of calibration option prices contains both Puts and Calls, and about 3\% for Puts only.

\subsubsection{Constraints on the model parameters}

In \cite{HalperinItkin2025ar}, the Marketron model was calibrated to market prices of the S\&P500 index using particle filtering, subject to constraints on model parameters inspired by the shape of the Marketron potential. However, our numerical experiments with calibrating the model to option prices revealed a significant obstacle. Since the model has many parameters, the least-square objective function used in calibration contains numerous local minima. Therefore, even when using a global solver, limited computational resources prevent us from identifying the global minimum. To address this challenge, we impose additional constraints on model parameters to ensure the solution remains financially meaningful.

Looking at the model definition in \eqref{Marketron_3D}, one can see that the SDE for the memory variable $y_t$ has a mean-reverting drift with the total mean-reversion level $\hat{y}$ equal to
\begin{equation}
\hat{y} = \frac{h(\theta_t)}{\mu} + \bar{y} - \frac{c}{\mu} V_M(x).
\end{equation}
This mean-reversion level depends on other stochastic drivers $\theta_t$ and $x_t$. In turn, the signal variable $\theta_t$ is also mean-reverting with the mean-reversion level equal to $\hat{\theta}$. Finally, the log-price variable $x_t$ could also be mean-reverting depending on the signs of $f(\theta_t) + \eta$ and $c y_t V'_M(x_t)$.

Without additional constraints, the drift of $x_t$ per unit $dt$ (which is approximately an annualized mean of log-returns) could be unrealistically high or low due to the behavior of $V'_M(x)$ and $V_M(x)$ at $x \to \pm \infty$, see the definition of these functions in \eqref{Langevin_potential_exact}. But what we actually would expect from the calibrated model is

\begin{itemize}
\item If $x \to \infty$, it implies that $V_M(x) > 0$. Then we want $\hat{y}$ to be finite (at least at some $T > t > t_0$), otherwise, the annualized drift of $x_t$ would produce unrealistic mean values of log-returns and unrealistic default rates.

\item When $x_t \to -\infty$, the model produces a default event. However, not all lower values of $x_t$ should give rise to default. Therefore, we need an additional condition to allow this behavior only for $x_t$ below some threshold $\hat{x}$. For higher (though still negative) values of $x_t$, we expect a mean-reverting behavior, so the annualized drift produces reasonable (close to known market) mean values of log-returns, and simultaneously, reasonable values of the default probabilities.
\end{itemize}

With this reasoning in mind, we impose the following additional constraints on the model parameters:
\begin{alignat}{2} \label{addConstr}
\bar{r} &> |f(\theta_0) + \eta - c V'_M(x_0) y_0|, &\qquad \bar{r} &> |f(\hat{\theta}) + \eta - c V'_M(\hat{x}) \hat{y}|.
\end{alignat}
Here $\bar{r}$ is a typical market value of the annualized mean of log-returns. In our numerical experiments, we set $\bar{r} = 0.02$ and $\hat{x} = x_0 - 2$.

Bear in mind that this problem does not appear in \cite{HalperinItkin2025ar} since particle filtering is performed at every moment of time $t$ using the first four corresponding market moments as a benchmark. However, when calibrating to option prices, we have the necessary market data only at the inception of the option contract and at maturity.

Also, the box constraints on some model parameters, e.g., $\bar{y}$ have been extended as now the role of $\bar{y}$ is played by $\hat{y}$, etc.

\subsubsection{Results} \label{results}

We begin with selecting the option maturity $T = 0.425$, corresponding to a 22-week period. Given the limited number of strikes meeting our criteria (typically 10 per date), we aggregate option data from multiple dates -- specifically, January 17 to February 15, 2017 -- yielding a total of 193 option prices for calibration. The calibration results are summarized in Table~\ref{resT0_425}.
\begin{table}[tbhp]
\begin{center}
\scalebox{0.87}{
\begin{tabular}{|l|r|r|r|r|r|r|r|}
\toprule
\rowcolor[rgb]{ .792,  .929,  .984} {\bf parameter} & $\bm \sigma$ & $\bm \sigma_y$ & $\bm \sigma_z$ & $\bm k$ & $\bm \mu$ & $\bm g$
& ${\bm \hat{\theta}}$ \\ \hline
value       & 0.3934 & 1.008 & 0.8912 & 2.7069 & 4.6154 & 0.3173 & 6.9242 \\ \hline
\rowcolor[rgb]{ .792,  .929,  .984} {\bf parameter} & $\bm c$ & $\bm b_1$ & $\bm b_2$ & ${\bm \bar{y}}$ & $\bm \gamma$ & $\bm y$
& $\bm \theta$ \\ \hline
value       & 0.8897 & 0.1220 & -0.0549 & 1.6208 & 1.1031 & -0.0589& 1.1007 \\ \hline
\bottomrule
\end{tabular}
}
\caption{Parameters of the model in \cref{Marketron_3D,defTanh} found by calibration to SPX options prices at $T = 0.425$ for data from Jan.17, 2017 to Feb.15, 2017.}
\label{resT0_425}
\end{center}
\end{table}
A comparison of the calibrated model parameters reveals that the volatility $\sigma_z$ and parameter $\hat{\theta}$ are relatively close to those obtained in \cite{HalperinItkin2025ar} from equity data calibration. The remaining parameters exhibit some differences. This divergence can be attributed to several key factors. First, the model specifications for $f(\theta)$ and $h(\theta)$ differ between \cite{HalperinItkin2025ar} and our work in \eqref{defTanh}. Specifically, the former uses a time-dependent formulation, while our model assumes constant parameters. Second, the initial values $y_0 = y, \theta_0 = \theta$ in \cite{HalperinItkin2025ar} were set to zero, while in our study, these are also calibration parameters. Finally, it is not obvious at all that both calibrated sets should be close enough despite that would be a nice feature.

We also examine the case $T = 0.041$ (2-weeks maturity). Due to the limited number of eligible strikes (as per our selection criteria), we again aggregate option data across multiple dates — January 17 to February 9, 2017 — resulting in 140 option prices for calibration. The corresponding results are presented in Table~\ref{resT0_041}. It can be seen that, except $\sigma_z, \mu, c$, thus found parameters are either close or of the same order of magnitude to the parameters reported in \cite{HalperinItkin2025ar} and obtained by calibrating the Marketron model to the SPX500 market data.
\begin{table}[tbhp]
\begin{center}
\scalebox{0.87}{
\begin{tabular}{|l|r|r|r|r|r|r|r|}
\toprule
\rowcolor[rgb]{ .792,  .929,  .984} {\bf parameter} & $\bm \sigma$ & $\bm \sigma_y$ & $\bm \sigma_z$ & $\bm k$ & $\bm \mu$ & $\bm g$
& ${\bm \hat{\theta}}$ \\ \hline
value       & 0.8950 & 0.1244 & 0.2004 & 1.8831 & 4.5869 & 0.3108 & 7.5284 \\ \hline
\rowcolor[rgb]{ .792,  .929,  .984} {\bf parameter} & $\bm c$ & $\bm b_1$ & $\bm b_2$ & $\bm \bar{y}$ & $\bm \gamma$ & $\bm y$
& $\bm \theta$ \\ \hline
value       & 1.1189 & 0.2455 & 1.1286 & 1.1148 & 5.4118 & -0.2356 & -0.2014 \\ \hline
\bottomrule
\end{tabular}
}
\caption{Parameters of the model in \cref{Marketron_3D,defTanh} found by calibration to SPX options prices at $T = 0.041$ for data from Jan.17, 2017 to Feb.9, 2017.}
\label{resT0_041}
\end{center}
\end{table}

\begin{figure}[!htb]
\captionsetup[subfloat]{captionskip=-10pt}
\begin{center}
\scalebox{0.83} {
\hspace*{-0.4in}
\subfloat[]{\includegraphics[width=0.6\textwidth]{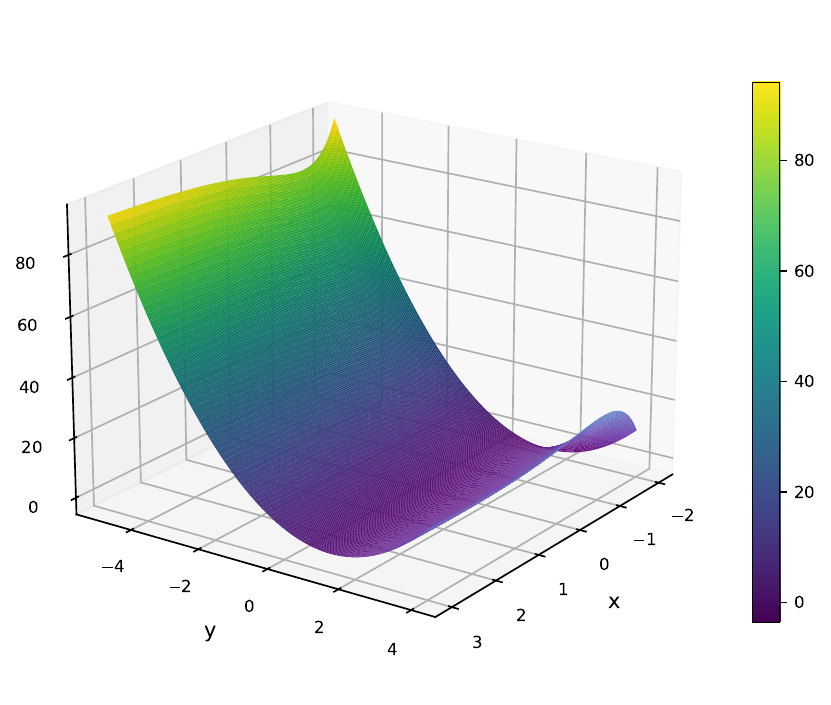}}
\subfloat[]{\includegraphics[width=0.6\textwidth]{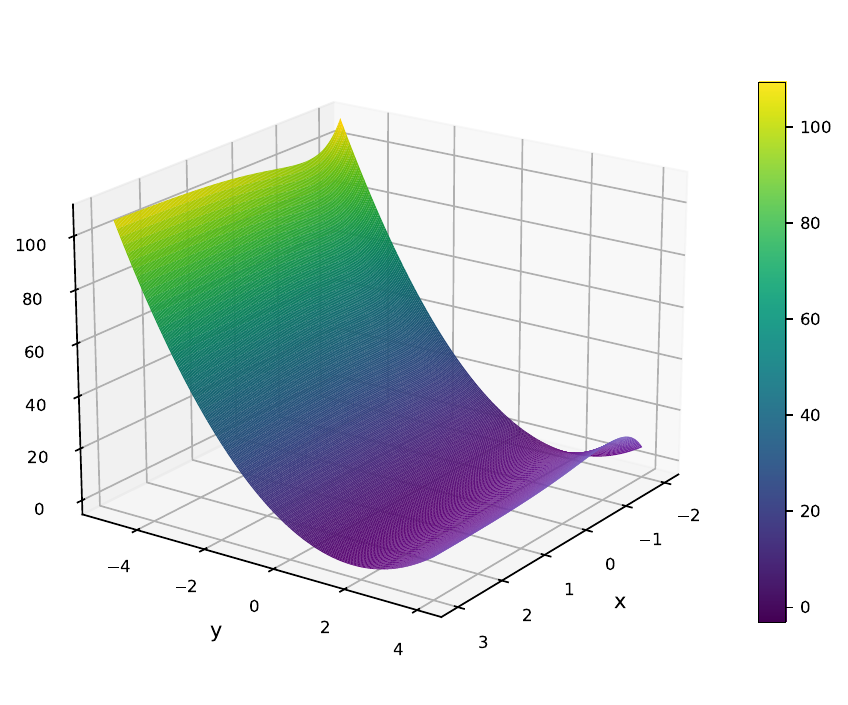}}
}
\end{center}
\caption{3D marketron potential $V(x,y)$ in \eqref{Langevin_potential_exact} computed with the model parameters found by calibration for a) $t=0.041$ year, b) $t=0.425$ years.}
\label{pot3Dcalib}
\end{figure}

Fig.~\ref{pot3Dcalib} shows a 3D marketron potential $V(x,y)$ in \eqref{Langevin_potential_exact} computed with the model parameters found by calibration for maturities $T=0.041$ and $T=0.425$ years.

It can be seen that the form of the potential differs from that in \cite{HalperinItkin2025ar}. Here, the potential represents just a single well in the $y$ space. A similar result was reported in \cite{NES}, where a comparable but simplified 1D model was used.

Examining the calibrated values of the risk-aversion parameter $\gamma$, we observe that $\gamma \gg 1$ for intermediate maturities. Values of $\gamma$ exceeding one have been consistently reported in the literature - see, for example, \cite{Bliss2004}. However, for $T = 0.041$, our calibration yields $\gamma = 1.1031$, i.e., the risk aversion parameter is close to one. While this value does not violate the condition $\gamma > 0$, intuitively it appears somewhat unusual. That said, the theoretical range of $\gamma$ may vary depending on the specific model being used.

\subsubsection{Implied distribution of log-returns from the options implied volatilities} \label{implDist}

As introduced earlier, this paper aims to develop a calibration approach for the Marketron model using market option data. Once calibrated, the next question we would be interested to answer is whether the Marketron model can simultaneously reproduce the statistical properties of the underlying asset's log-returns.

The problem of extracting implied probability distributions of underlying assets from option market data has drawn attention in the literature, see e.g., \cite{Sestanovic2018} and references therein. As stated by the authors, moments of future prices and returns are not observable, but it is possible to measure them indirectly. A set of option prices with the same maturity but with different strikes can be used to extract the implied probability distribution of the underlying asset at the expiration date and its moments. The aim is to obtain market expectations from options and to investigate which non-structural model for estimating implied probability distributions gives the best fit.

In particular, they consider three models: mixture of two log-normals (MLN), Edgeworth expansions, and Shimko's model (representatives of parametric, semiparametric, and nonparametric approaches respectively). The best fit model is used to describe moments of the implied probability distribution. The sample covers one-year data for DAX index options. The obtained results demonstrate that all models give better short-term forecasts. For 1M and 2M options, the skew could be negative for one model and positive for another while they could be of the same sign depending on the inception date.

This approach is conceptually aligned with several studies that propose models capable of simultaneous calibration to both S\&P500 option prices and VIX derivatives, e.g., see \cite{Guyon2023, AbiJaber2023, Cuchiero2024} among others. Typically this is achieved by introducing various additional stochastic factors to enhance flexibility of the model. However, in \cite{NES} it is demonstrated that a simplified one-dimensional version of the Marketron model can be successfully calibrated to market option prices (either calls or puts). From this calibration, the annualized moments of the implied risk-neutral distribution are computed, yielding values consistent with observed market data. Notably, the model in \cite{NES} is also based on a mixture of Gaussian distributions.

In a similar vein, this section leverages the calibrated model parameters to compute various statistics by solving \eqref{Marketron_3D} via Monte Carlo simulation, following the methodology outlined in \cite{HalperinItkin2025ar}. This approach provides deeper insights into the dynamic behavior of the model. For instance, Table~\ref{Stats} presents the annualized first four moments (the last two are also normalized) of the log-returns distribution computed using parameters calibrated at $T = 0.041$. The time horizons displayed in this table are approximately 2 weeks, 1 month, 3 months, 6 months, 1 and 3 years. Since for calibration, options with maturity $T = 2$ weeks have been used in this experiment, it would be naive to expect that predicted log-returns could represent the market at longer maturities; therefore, we restrict the time horizon to 3 years. In turn, Table~\ref{Stats2} presents statistical results for the longer maturity $T=0.425$.

The analysis reveals that for the short horizons (less than 6 months for 2w options, and less than 1 yr for 6m options) the log-returns exhibit a very small skewness and kurtosis which could be both negative and positive. For longer horizons, we observe a positive skewness and kurtosis of the time-series obtained by simulation. In contrast, the results in \cite{HalperinItkin2025ar} demonstrate negative skewness of log-returns for all horizons and positive kurtosis.
\begin{table}[!htb]
\centering
\begin{tabular}{|r|r|r|r|r|}
\toprule
\rowcolor[rgb]{ .792,  .929,  .984} \multicolumn{1}{|c|}{\textbf{horizon, yrs}} & \multicolumn{1}{c|}{\textbf{mean}} & \multicolumn{1}{c|}{\textbf{volatility}} & \multicolumn{1}{c|}{\textbf{skewness}} & \multicolumn{1}{c|}{\textbf{kurtosis}} \\ \hline
0.0397 & -0.1034 & 0.6331 & 0.0003 & -0.0001 \\ \hline
0.0833 & -0.1085 & 0.7734 & 0.0012 & 0.0004 \\ \hline
0.25 & -0.1086 & 0.8340 & 0.0112 & 0.0006 \\ \hline
0.50 & -0.0981 & 0.8022 & 0.0546 & -0.0082 \\ \hline
1.00 & 0.0855 & 0.7423 & 0.1688 & -0.0074 \\ \hline
2.00 & 0.2038 & 0.6400 & 0.4535 & 0.2740 \\ \hline
2.80 & 0.2028 & 0.5932 & 0.5852 & 0.7774 \\ \hline
3.00 & 0.2005 & 0.5895 & 0.5482 & 1.0588 \\
\bottomrule
\end{tabular}%
\caption{Annualized statistics of log-returns computed by using the model \eqref{Marketron_3D} with the model parameters found by calibration to SPX options prices at $T = 0.041$.}
\label{Stats}%
\end{table}%
\begin{table}[!htb]
\centering
\begin{tabular}{|r|r|r|r|r|}
\toprule
\rowcolor[rgb]{ .792,  .929,  .984} \multicolumn{1}{|c|}{\textbf{horizon, yrs}} & \multicolumn{1}{c|}{\textbf{mean}} & \multicolumn{1}{c|}{\textbf{volatility}} & \multicolumn{1}{c|}{\textbf{skewness}} & \multicolumn{1}{c|}{\textbf{kurtosis}} \\ \hline
0.0397 & -0.0136 & 0.2781 & -0.0006 & -0.0004 \\ \hline
0.0833 & -0.0114 & 0.3403 & 0.0006 & -0.0004 \\ \hline
0.25 & 0.0505 & 0.3693 & -0.0011 & 0.0030 \\ \hline
0.50 & 0.2099 & 0.3606 & 0.0058 & 0.0043 \\ \hline
1.00& 0.3558 & 0.3333 & 0.0533 & 0.0175 \\ \hline
2.00& 0.3793 & 0.2991 & 0.1807 & 0.0730 \\ \hline
2.80& 0.3587 & 0.2831 & 0.2914 & 0.1473 \\ \hline
3.00 & 0.3548 & 0.2812 & 0.3147 & 0.1686 \\
\bottomrule
\end{tabular}%
\caption{Annualized statistics of log-returns computed by using the model \eqref{Marketron_3D} with the model parameters found by calibration to SPX options prices at $T = 0.425$.}
\label{Stats2}%
\end{table}

Thus, examining the S\&P500 time-series generated by simulations of the Marketron model - calibrated a) to S\&P500 data in \cite{HalperinItkin2025ar} and b) to SPX options market data - reveals a difference in the sign of skewness. To underline, in this paper we focus on the fixed period from 2017 to 2020, whereas \cite{HalperinItkin2025ar} calibrates the model over a 25-year span. Our findings show that the Marketron model successfully replicates the sign of kurtosis in both short- and long-term scenarios. However, the skewness exhibits an opposite sign between the two cases, raising the question of which sign is correct.

Thinking of that, first note that the negative skewness of the S\&P500 time-series has been well-documented in prior literature, such as \cite{NeubergerPayne2019}, where the authors use proxy techniques on U.S. stock index returns and demonstrate persistently negative skewness across time horizons, from monthly to multi-year periods. Conversely, \cite{Xu2024} shows that from 2017 to 2021, markets exhibited an inverted pattern - taking the "escalator down and the elevator up" - resulting in positive realized skewness for SPX during this period.

Additionally, \cite{Kownatzki2025} explores the use of intraday market dynamics to predict major turning points by analyzing skewness, kurtosis, and the Hurst exponent. The study examines minute-by-minute S\&P500 (SPX) and NASDAQ100 (NDX) data, particularly during the COVID-19 pandemic and the Great Financial Crisis (GFC). The results indicate that before market tops, skewness becomes more negative, kurtosis rises, and the Hurst exponent trends upward. But the exact opposite trends were observed just before a market bottom.

In \cite{Farago2023}, it is argued that multiplicative compounding at long horizons induces strong-to-extreme positive skewness in stock returns, with the effect's magnitude primarily determined by single-period volatility. They reference \cite{Bessembinder2018}, who empirically demonstrates that long-run compound stock returns behave very differently from short-run (monthly or annual) returns. Through simulation exercises, Bessembinder illustrates how compounding induces strong positive skewness in multiperiod returns - even when single-period returns are symmetric. While \cite{Bessembinder2018} primarily focuses on individual stocks, his simulation results suggest that this skew-inducing effect of compounding should also appear in aggregate returns, albeit to a lesser extent. Also, \cite{Farago2023} notes that in stark contrast, \cite{NeubergerPayne2019} present a contrasting view, arguing that long-run aggregate stock returns are substantially negatively skewed.

Therefore, this problem requires a more detailed analysis which is left for the future research. Anyway, it would be interesting to perform even a simple validation of the moments presented in Table~\ref{Stats2}. Here is how we approach this.

Recall that in \cite{HalperinItkin2025ar}, we calibrated the Marketron model to S\&P 500 daily log-prices from January 2000 to October 2024. Using this data, we extract the time-series around 2017 and compute statistical moments (mean, volatility, skewness, and kurtosis) of the historical distribution using rolling windows of 0.5 and 1 year for each day from January 1, 2017, to February 9, 2018. We then compare these results with those in Table~\ref{Stats2}. Bear in mind that comparing short-maturity option statistics is less meaningful, as stock data over such brief periods yields unreliable statistics. The comparison is presented in Table~\ref{comp} and briefly summarized below
\begin{table}[htbp]
  \centering
\begin{tabular}{|r|r|r|r|r|}
\toprule
\rowcolor[rgb]{ .792,  .929,  .984} \multicolumn{1}{|l|}{\textbf{horison, yrs}} & \multicolumn{1}{l|}{\textbf{mean}} & \multicolumn{1}{l|}{\textbf{volatility}} & \multicolumn{1}{l|}{\textbf{skew}} & \multicolumn{1}{l|}{\textbf{kurtosis}} \\
\midrule
\rowcolor[rgb]{ 1,  1,  0} \multicolumn{5}{|c|}{\textit{Simulation using the Marketron calibrated to option data}} \\
\hline
0.5   & 0.2099 & 0.3606 & 0.0058 & 0.0043 \\
\hline
1.0   & 0.3558 & 0.3333 & 0.0533 & 0.0175 \\
\hline
\rowcolor[rgb]{ 1,  1,  0} \multicolumn{5}{|c|}{Using the SPX time-series around 2017} \\
\hline
0.5   & 0.1945 & 0.0756 & 0.0314 & 0.0103 \\
\hline
1.0   & 0.2055 & 0.1083 & -0.0293 & 0.0175 \\
\bottomrule
\end{tabular}
\caption{Annualized statistics of log-returns computed by using the model \eqref{Marketron_3D} with the model parameters found by calibration to SPX options prices at $T = 0.425$ and SPX time-series.}
\label{comp}%
\end{table}

For the 0.5-year horizon:
\begin{itemize}
\item The simulated mean is relatively close to the time-series mean.
\item Skewness and kurtosis share the same sign, though the simulated values are smaller in magnitude (all values remain quite small).
\item Simulated volatility is significantly higher than that of the time-series.
\end{itemize}
For the 1-year horizon:
\begin{itemize}
\item Kurtosis aligns closely between the two datasets.
\item Time-series volatility remains much lower than the simulated values.
\item The simulated mean is higher, while the time-series mean shows little change.
\item The time-series skew becomes slightly negative, whereas the simulated skew remains positive (though both are small).
\end{itemize}

Thus, the comparison between the simulated moments in Table~\ref{Stats2} and time-series statistics in Table~\ref{comp} reveals both consistencies and discrepancies. The main discrepancy is about volatility - the model consistently overestimates volatility compared to the historical data, suggesting that the Marketron calibration may either overstate market fluctuations or miss certain dampening effects present in real price movements.  Further investigation could explore whether these differences stem from calibration choices, model structure, or inherent limitations in capturing rare events or regime shifts in the S\&P 500.

Our current view on this discrepancy is as follows. In 2017, the S\&P 500 exhibited remarkably low volatility, with implied volatility (IV) typically exceeding historical volatility (HV). The annualized daily volatility that year was only around 6.6\% - roughly one-third of the historical average. In fact, 2017 is often regarded as the least volatile year for U.S. equities in decades. The S\&P 500’s maximum peak-to-trough drawdown was a mere -2.8\%, ranking among the smallest intra-year losses in history. Despite events such as hurricanes, wildfires, and geopolitical tensions, market conditions remained unusually calm. At the same time, the IV, though a forward-looking measure, tended to be upward-biased. This divergence between the IV and HV presented potential trading opportunities in the options market.

Given these observations, we conclude that the Marketron model fails to solve the joint calibration problem, unlike the approaches in \cite{Guyon2023,AbiJaber2023,Cuchiero2024}, despite incorporating memory effects. Potential reasons for this limitation include: i) the model assumes uncorrelated BMs, whereas nonlinear drifts may be insufficient to replicate the necessary correlations; and ii) the framework in \cite{Guyon2023} requires a path-dependent local volatility function, a feature absent (at  least, explicitly) in our model.

\begin{figure}[!htb]
\begin{center}
\includegraphics[width=0.7\textwidth]{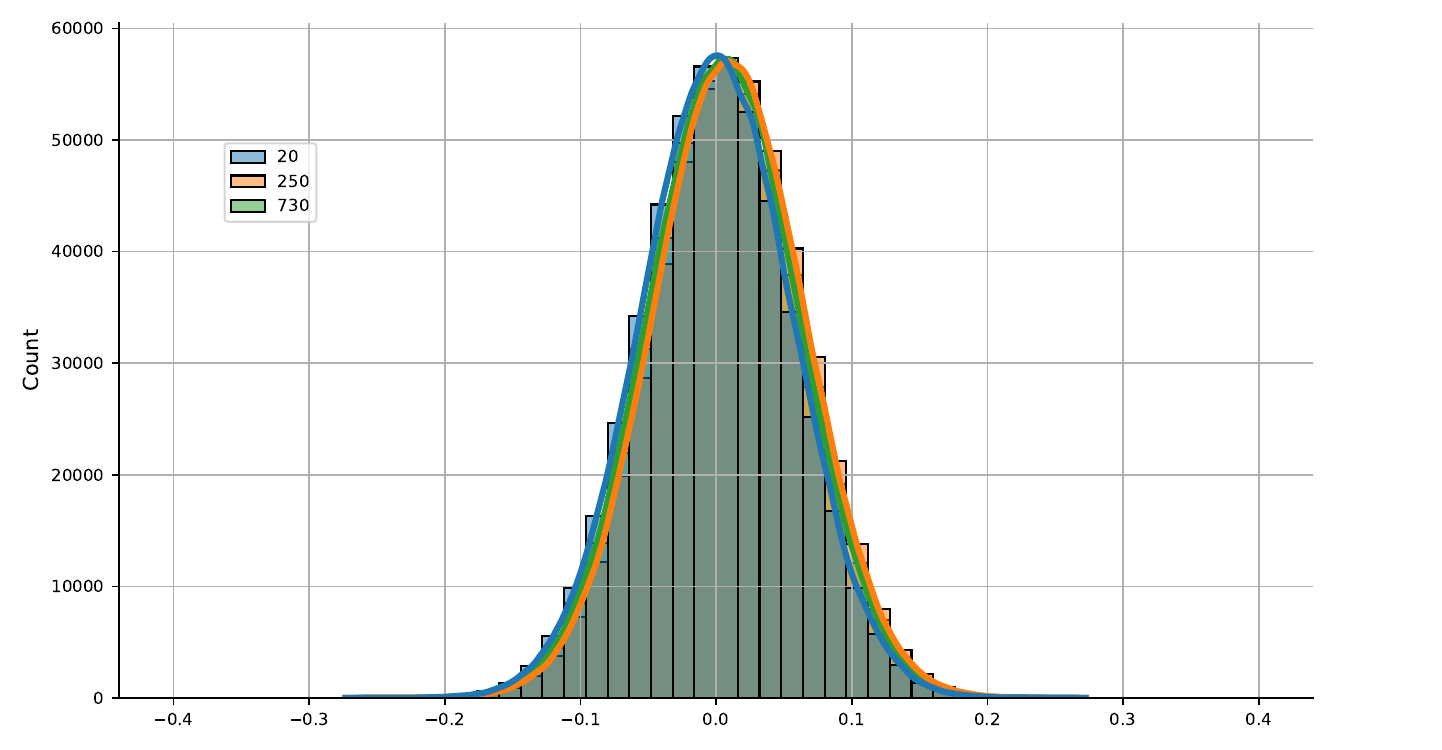}
\end{center}
\vspace{-2em}
\caption{The distributions of the log-returns at $t=20, 250, 730$ business days, obtained by simulation with the model parameters found by calibration to the option data with $T=0.425$.}
\label{dist0_425}
\end{figure}

Fig.~\ref{dist0_425} presents a distribution of the log-returns, obtained by simulation with the model parameters found by calibration to the option data with $T=0.425$, at three moment of times measured in weeks.

\paragraph{Other characteristics.} The {\bf default probability} obtained in our simulations for $T=0.041$ is 345 basis points. This value is a bit high but approximately of the same order of magnitude as those reported in \cite{HalperinItkin2025ar}. To recall, the default probability is partly regulated by additional constraints imposed in \eqref{addConstr}. Normally, we expect $p_d$ at short maturities to be smaller, while our calculations show the opposite. This can be explained by the fact that for longer maturities $\hat{x}$ has to be moved further left since $x_t$ has more time (and thus higher probability) to reach this threshold during the life of the option. When we keep this threshold close to $x_0$, it effectively reverts the defaulted paths back, thus decreasing the default probability $p_d$. This behavior also should be investigated in more detail in the future work.

The {\bf Hurst exponent} measured at $T=0.425$ is 0.306, while at $T=0.041$ it is 0.294, which can be compared with the value 0.61 in \cite{HalperinItkin2025ar}. It is worth noting that while rough volatility models typically generate low Hurst exponents at short maturities in option pricing contexts, our results represent the Hurst exponent of log-return time-series simulated using the Marketron model calibrated to option prices at short maturities. Nevertheless, the Hurst exponent obtained for our both maturities is close to values reported in the literature \cite{Livieri2018}\footnote{In \cite{Livieri2018}, the value of the Hurst exponent obtained by using high frequency volatility estimations from historical price data has been revisited by studying implied volatility based approximations of the spot volatility. Using at-the-money options on the S\&P500 index with short maturity, the authors confirm that volatility is rough, and the Hurst parameter is of order 0.3, i.e., slightly larger than that usually obtained from historical data.}, but is almost two times less than those reported in \cite{HalperinItkin2025ar}, which were obtained by calibrating the model to equity time-series data, and they partly justify some weak volatility clustering at the short time-end. Also, the accuracy of computing the Hurst exponent is sensitive to the range of time lags considered for the given time-series.

\subsubsection{The market price of risk} \label{MPR}

\begin{figure}[!htb]
\begin{center}
\includegraphics[width=0.8\textwidth]{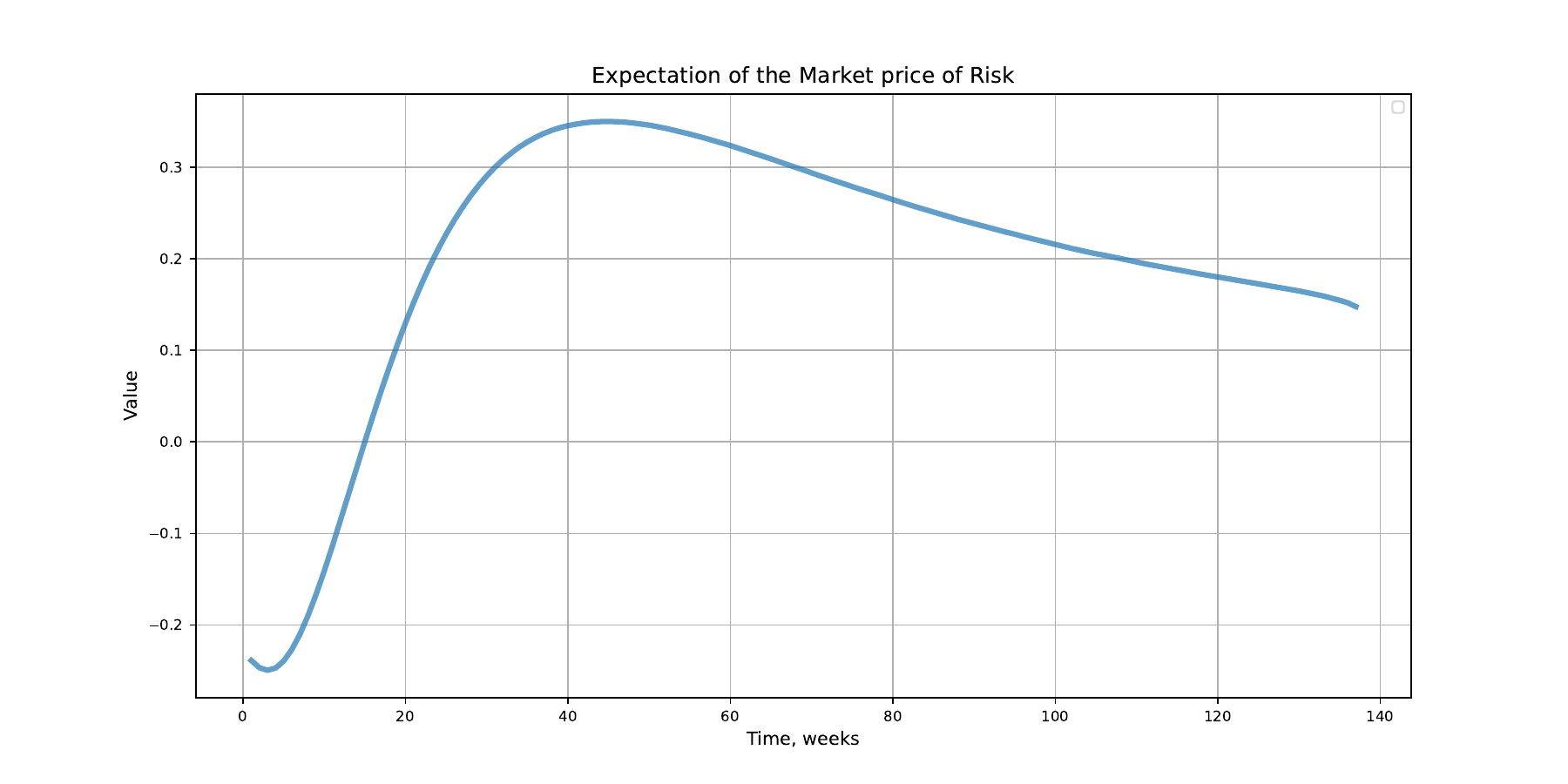}
\end{center}
\vspace{-2em}
\caption{Expectation of the MPR computed in MC simulation of the Marketron model with parameters found by calibration to the option market data at $T=0.041$.}
\label{mpr0_041}
\end{figure}
\begin{figure}[!htb]
\begin{center}
\includegraphics[width=0.8\textwidth]{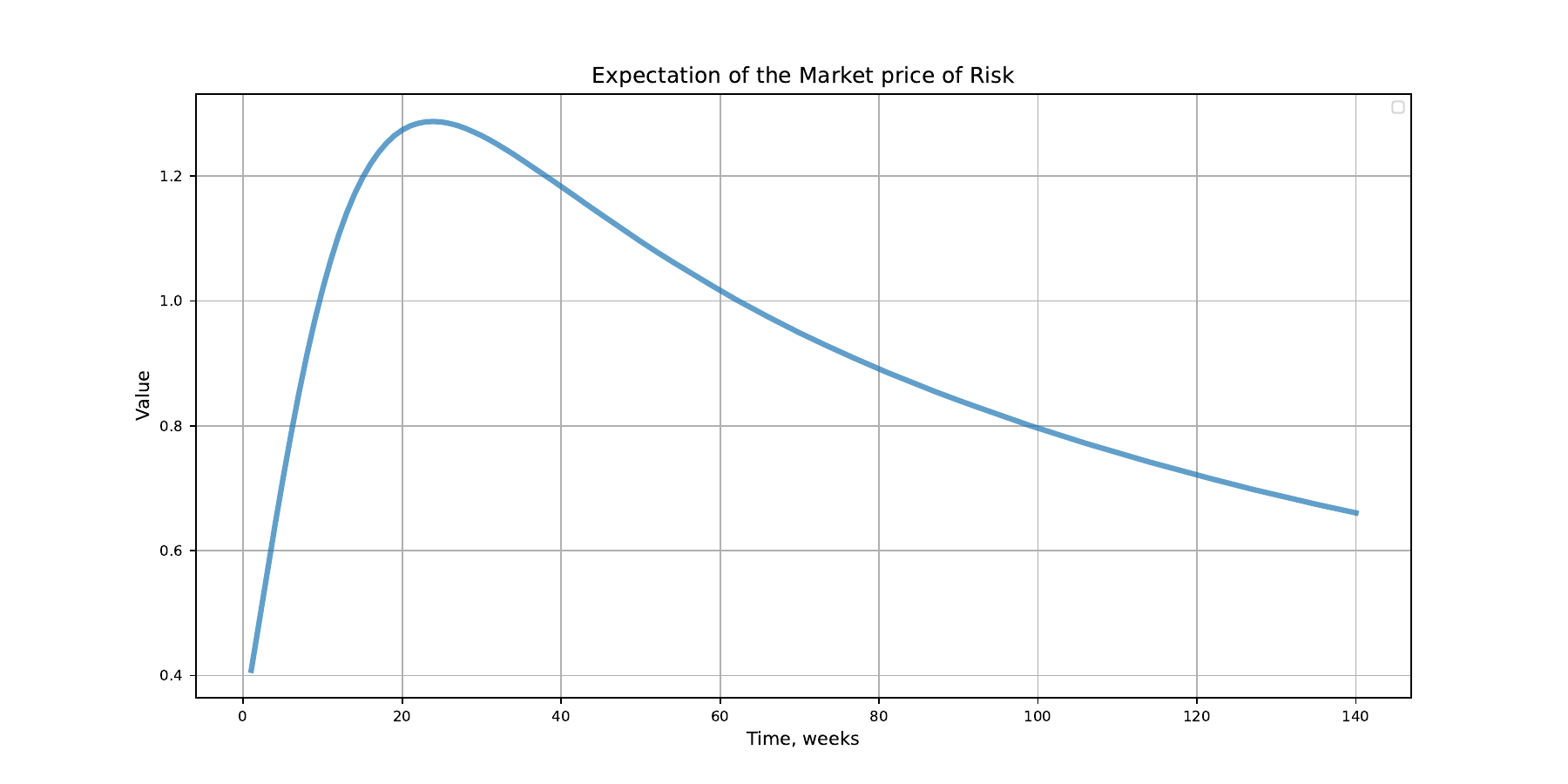}
\end{center}
\vspace{-2em}
\caption{Expectation of the MPR computed in MC simulation of the Marketron model with parameters found by calibration to the option market data at $T=0.425$.}
\label{mpr0_425}
\end{figure}
When performing the simulation, we also computed the market price of risk (MPR). Since the MPR in our model is stochastic due to its dependence on the model's state variables, we calculated it along each simulated path and then took the expectation. The resulting time series of the MPR are shown in Fig.~\ref{mpr0_041} for $T = 0.041$ and in Fig.~\ref{mpr0_425} for $T = 0.425$. When computing the MPR the defaulted paths were excluded.

It can be seen that the MPR for the short $T$ could be negative which aligns with what is often reported in the literature. At the same time, the analysis in, e.g., \cite{Mikosch2004} provides statistical evidence that the expected return of the S\&P500 index as well as the market price of risk  vary through time both in size and sign. In particular, the periods of negative (positive) expected return and market price of risk coincide with the bear (bull) markets of the index as defined in the literature.

\begin{figure}[!htb]
\begin{center}
\includegraphics[width=0.9\textwidth]{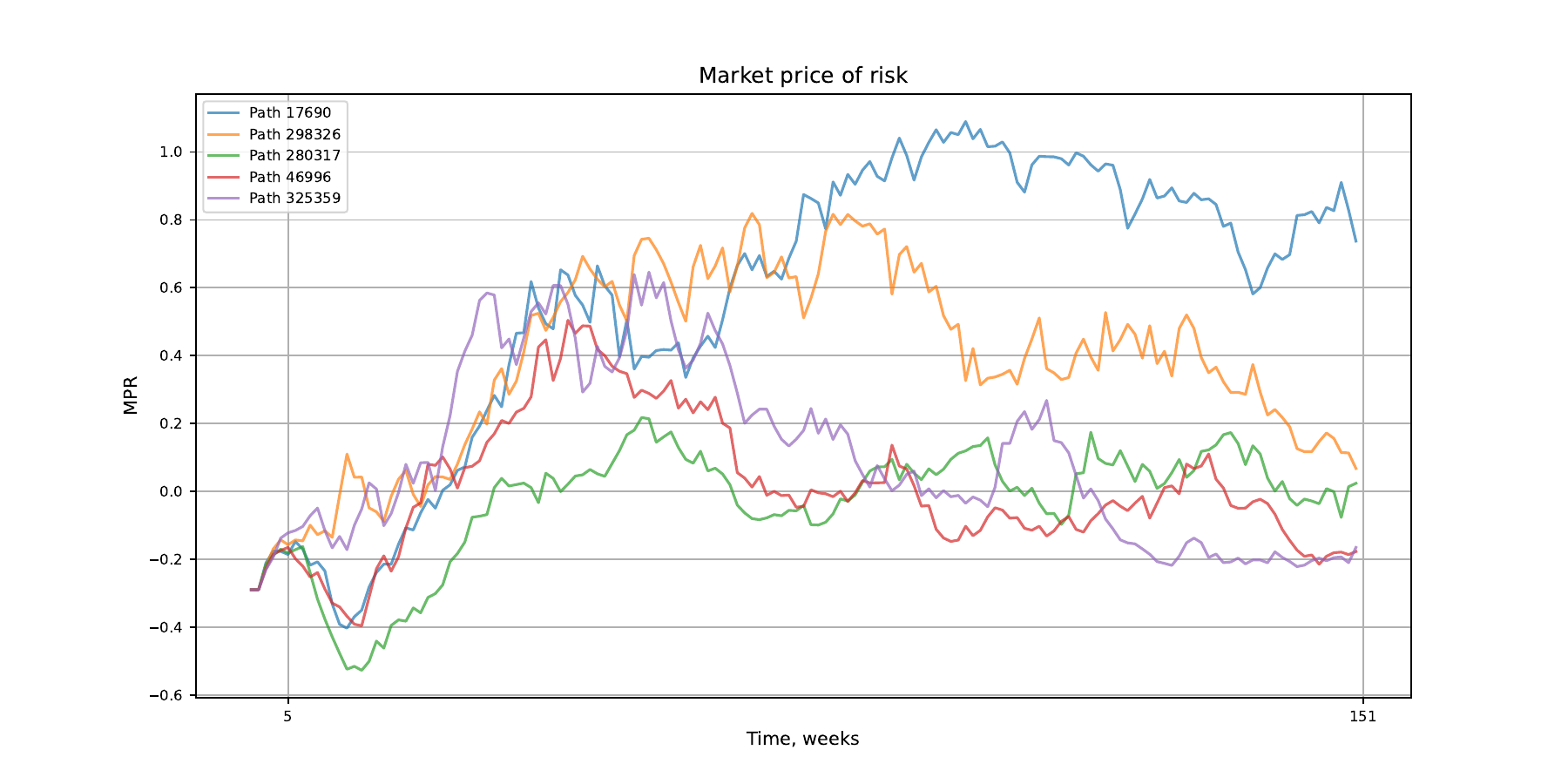}
\end{center}
\vspace{-2em}
\caption{The market price of risk along random paths as a function of the time computed in MC simulation of the Marketron model with parameters found by calibration to the option market data at $T=0.041$.}
\label{mprPaths0_041}
\end{figure}

We also plot the market price of risk along random paths as a function of the time obtained in simulation, which are shown in Fig.~\ref{mprPaths0_041}

\section{Discussion}

\paragraph{Calibration of the Marketron model to market option data.} To recall, the concept of this paper is as follows. In \cite{HalperinItkin2025ar}, we calibrated the Marketron model to SPX500 time-series data and examined whether the calibrated model could reproduce key stylized market behaviors and measures. This paper follows the same approach but calibrates the model to SPX options instead. Since this requires working under the risk-neutral measure, we employ indifference pricing combined with an investor utility framework to construct it. While solving the resulting HJB equation presents significant computational challenges, we have developed a novel method that enables calibration within reasonable timeframes on standard laptop hardware.

To briefly summarize, the key steps of our method are as follows:

\begin{itemize}
\item Using indifference pricing and following \cite{GrasselliHurd2007}, we derive a Hamilton-Jacobi-Bellman (HJB) equation for the risk-neutral option price.

\item Solving the maximization problem yields a nonlinear PDE for the option price, featuring quadratic nonlinearities in the first derivative of the price. Additionally, the coefficients of this PDE are highly nonlinear functions of the model parameters.

\item While this PDE can be solved via MC or FD methods, these approaches are computationally expensive. Instead, we propose two alternative methods, with operator splitting, \cite{LanserVerwer,ItkinBook}, proving superior. This technique decomposes the original unsteady 3D equation into a sequence of 1D equations - some linear with nonlinear coefficients, others fully nonlinear.

\item We demonstrate that the nonlinear equations can be transformed into linear counterparts using the Cole-Hopf transformation.

\item Since all stochastic terms are Gaussian (though the model itself is non-Gaussian due to nonlinear drifts), the Green’s function for each PDE is known. Applying a generalized Duhamel's principle, \cite{Itkin2024jd,Hunter2014}, we reformulate each PDE as a Volterra integral equation of the second kind with a known kernel.

\item To solve these 1D integral equations efficiently, we employ a Gaussian RBF method. Being meshless, this approach drastically reduces the required number of nodes compared to traditional PDE solvers (e.g., FD methods), which demand dense grids.

\item Leveraging the Gaussian structure of the RBFs, we develop efficient closed-form approximations for the nonlinear coefficients, enabling closed-form evaluation of the integrals in the Volterra equations (see \cref{appMatrixRBF}).
    
\item Overall, the above steps reduce the full problem to solving a number of linear systems for RBF coefficients with multiple right-hand parts representing different strikes.

\item Finally, we implement a parallelized calibration algorithm, exploiting advanced features of Python and Polars.
\end{itemize}

This comprehensive framework is new, solves the problem of calibration of the Marketron model to market option data,  and represents a significant methodological advancement, enabling efficient model calibration to option market data while maintaining analytical tractability.

Despite its advantages, including computational efficiency, our model requires the calibration of 15 parameters, which may hinder its speed for practical calibration purposes. To address this, domain knowledge (where available) could help reducing the parameter space to a more manageable size. Alternatively, one could employ a feedforward artificial neural network (ANN) or a physics-informed neural network (PINN). These networks can be trained on in-sample option market data to determine their weights. Once trained, standard least-squares calibration can be performed using the neural network as a fast surrogate for the model pricer (for details, see, e.g., \cite{Horvath2019,Kees2019,ItkinDL2020} and references therein). We may explore this approach in future work.

\paragraph{Replication of the underlying implied distribution.} Beyond the methodological contributions, sections \cref{results,implDist,MPR} mirror the analytical framework from \cite{HalperinItkin2025ar}, featuring comparable plots and tables. We are trying to address a popular question in current quantitative finance literature: can a single model simultaneously predict equity returns, option smile dynamics, and perhaps volatility indices like VIX? This challenge has attracted considerable attention within last decade. Such unified modeling approaches hold a particular value for traders seeking to extract actionable information from options markets for equity trading strategies.

Our results indicate that the Marketron model, when calibrated to options under the risk-neutral measure, to some extent retains predictive power for underlying index characteristics when applied to equity markets in the real measure. In more detail, our tests sought to capture moments of the S\&P 500 historical distribution around 2017 through calibration of the Marketron model to options data with $T=0.425$ from the same period. This calibration revealed that the Marketron model fails to solve the joint calibration problem, unlike the approaches in \cite{Guyon2023,AbiJaber2023,Cuchiero2024}, despite incorporating memory effects. By "fails," we mean that while some moments (mean, skew, kurtosis) are replicated relatively well, the computed volatility aligns more closely with the implied volatility of options rather than with the historical volatility of the index.

Potential reasons for this limitation include:
\begin{itemize}
\item The model assumes uncorrelated Brownian motions, whereas nonlinear drifts may be insufficient to replicate the necessary correlations. 
\item The framework in \cite{Guyon2023} requires a path-dependent local volatility function—a feature that is absent (at least explicitly) in our model
\end{itemize}
The first point can be easily addressed since it implies adding another step to our splitting method which handles a mixed derivatives term. This can be done as explained in \cite{ItkinBook}. However, the second point is more challenged. In any case, our results represent preliminary findings, and substantial further investigation is warranted. These capabilities could prove valuable for trading strategies that leverage cross-market information flow between options and equity markets.

To underline, our calibration incorporates nonlinear constraints on the Marketron potential's functional form. These constraints may potentially:
\begin{itemize}
\item Exclude the true global minimum from the feasible parameter space, or
\item Overly restrict the solution domain, limiting the model's flexibility.
\end{itemize}
This constrained optimization framework could consequently affect both the calibration quality and the resulting parameter estimates.

\paragraph{General notes about the model.} We present an alternative calibration methodology for the Marketron model originally introduced in \cite{HalperinItkin2025ar}. While the original study employed S\&P500 log-returns, our approach utilizes SPX option prices (both Puts and Calls) as calibration instruments. Our results reveal differences in the calibrated parameters between the two approaches, particularly substantial for short maturities.

As previously noted, the model specifications for $f(\theta)$ and $h(\theta)$ differ between \cite{HalperinItkin2025ar} and our formulation in \eqref{defTanh}, with the former adopting a time-dependent structure while we assume constant parameters. These structural differences notwithstanding, the observed parameter variations suggest that the Marketron model's extensive parameter space (comprising 15 parameters in our implementation) may be overly complex, creating numerous local minima in the least-squares objective function. Consequently, despite employing global optimization techniques, our calibration procedure frequently converges to suboptimal solutions, yielding a modest accuracy of approximately 5\% -- a limitation that could be overcome through substantially greater computational investment.

An important innovation of our approach is the explicit evaluation of the market price of risk, as detailed in \cref{mpr}. We show that the market price of risk component associated with the first Brownian motion, $W^{(x)}_t$, takes the same form as in a complete market. However, due to the definition of $\mu_x$ in \eqref{Marketron_3D}, this component depends on the parameters of the Marketron potential -- a direct consequence of the model's nonlinear drift. Specifically, the market price of risk is influenced not only by the potential's parameters but also by the initial states $(x, y, \theta)$. This state-dependent behavior aligns with findings in prior work, such as \cite{DaiSingleton:99,Sbuelz2012,Mijatovic2013}.

Another critical aspect to emphasize is the simulation of memory effects. As discussed in \cite{HalperinItkin2025ar}, it is well-documented that return distributions implied by stock prices or options exhibit fat tails consistent with power laws. To model this phenomenon, the concept of rough volatility has been introduced for both equity and options markets. Various studies, such as \cite{GatheralJaissonRos2014}, demonstrate that historical volatility time series often display roughness significantly less that of standard BM, instead aligning more closely with fractional Brownian motion (fBM) -- see the comprehensive references at \cite{RVsite}. While fBM is a self-similar process capable of generating either long memory or rough paths, it cannot produce both simultaneously.

Over the past decade, rough volatility models have garnered significant attention. These models typically incorporate memory dependence in sample paths via fractional noise while retaining a deterministic drift. In contrast, our work demonstrates that memory effects can be embedded directly into the stochastic nonlinear drift of the Marketron model, even when the diffusion term is driven by a standard BM. Through calibration to options market data, we show that the Marketron model reproduces path roughness, quantified by the simulated Hurst exponent, without relying on fractional processes. This approach thus offers a novel pathway for integrating memory effects into the dynamics of financial assets.

Strictly speaking, the same should be said about jumps. The traditional approach to modeling jumps in asset market prices is through the stochastic component, where pure diffusion processes are replaced by \LY processes while retaining a deterministic drift. In contrast, the Marketron model retains a diffusion process but achieves jumps between states through "instantons" -- solutions of the dynamic equations that mediate sharp transitions between metastable market states. These instantons arise from the model's nonlinear drift and the specific structure of the Marketron potential (but not from the stochastic noise) of the SDE, as detailed in \cite{HalperinItkin2025ar}. This analogy could serve as a promising direction for future research. It is also worth mentioning that, as reported in \cite{Cuchiero2024}, the joint calibration problem can be solved using a stochastic volatility model where the dynamics of the volatility are described by a linear function of the (time extended) signature of a primary process which is supposed to be a polynomial diffusion, but also without adding jumps and rough volatility.

\section*{Acknowledgments}

We are grateful to Michael Isichenko for useful comments.

\section*{Disclaimer}

Opinions expressed here are author's own, and do not represent views of their employers. A standard disclaimer applies.

\printbibliography[title={References}]

\clearpage

\appendix
\appendixpage
\appendix
\numberwithin{equation}{section}
\setcounter{equation}{0}

\section{Explicit representation of $\calK(\tau_m, x, y, \theta; \calC)$}  \label{appMatrixRBF}

By definition in \eqref{trapezoid}, we have
\begin{align}
\calK(\tau_m, x, y, \theta; \calC) &= \int_{-\infty}^\infty \int_{-\infty}^\infty \int_{-\infty}^\infty \bar{\Phi}(\tau_m, \bm{p}; \calC) \, G(\Delta \tau_m; x,y,\theta| \bm{p}) d\bm{p}, \qquad \bm{p} = \{ \xi, \eta, \zeta \}, \quad d\bm{p} = d\xi d\eta d\zeta,
\end{align}
\noindent where $\bar{\Phi}(\tau_m, \bm{p}; \calC)$ was introduced in \eqref{Volterra} and the Green's function $G(\tau; x,y,\theta | \bm{p})$ was defined in \eqref{Green}. According to \eqref{Volterra}, $\bar{\Phi}(\tau_m, \bm{p}; \calC)$ contains terms that are linear in $\calC$, quadratic in $\calC$, or independent of $\calC$. Below, we derive closed-form expressions for all these terms.

Following \cite{HalperinItkin2025ar}, we assume that $c(t) = c$ is a constant to be determined by calibration. Consequently, the only terms that depend on time $\tau$ are the Green's function and $\bar{\Phi}(\tau_m, \bm{p}; \calC)$, the latter only through the RBF coefficients $c_{i,kjl}$ which vary with time $\tau_i$, and the function $f(\theta)$ as defined in \eqref{fhparam}. Furthermore, since we use a uniform time grid in $\tau$, the Green's function $G(\Delta \tau; x,y,\theta| \bm{p})$ remains the same for any time $\tau_i$ and thus needs to be computed only once.

\subsection{The linear part of function $\calK(\tau_m, x, y, \theta; \calC)$} \label{appLinear}

The linear in $\calC$ part of $\calK(\tau_i, x, y, \theta; \calC)$ can be written in the form
\begin{align} \label{Upsilon}
\Upsilon(\tau_i, x, y, \theta) = \sum_{k,j,l} c_{i,kjl} \int_{-\infty}^\infty \int_{-\infty}^\infty \int_{-\infty}^\infty \bar{\Psi}_1&(\tau_i, \bm{p}) G(\Delta \tau; x,y,\theta| \bm{p}) d\bm{p}.
\end{align}
To compute it, we need to substitute three components into the 3D integral on the right-hands part of \eqref{Upsilon}:
\begin{itemize}
\item The Green's function from \eqref{Green};
\item The function $\bar{\Phi}_1(\tau_i, \bm{p})$ from \eqref{VolterraRBF} given by
\begin{equation}
\bar{\Psi}_1(\tau_i, \bm{p}) = \bar{\eta} \calC_x(\bm{p}) + \mu_y(\tau_i, \bm{p}) \calC_y(\bm{p}) + \mu_\theta(\theta) \calC_\theta(\bm{p}) - r \calC(\bm{p});
\end{equation}
\item The RBF representation of $\mathcal{C}(\bm{p})$ using the Gaussian RBF defined in \eqref{gauss}
\begin{equation}
\calC(\bm{p}) = e^{-\varepsilon \left[ (\xi-x_k)^2 + (\eta-y_j)^2 + (\zeta - \theta_l)^2\right]},  \qquad \bm{p} = \{\xi, \eta, \zeta \}.
\end{equation}
\end{itemize}

Since $\bar{\Phi}_1(\tau_i, \bm{p})$ consists of three terms, we denote them as $\calA_m$, where $m \in [1,3]$. This allows us to express $\Upsilon(\tau_i, x, y, \theta)$ in the form
\begin{align} \label{upsA}
\Upsilon(\tau_i, x, y, \theta)  &= \sum_{m=1}^3 \int_{-\infty}^\infty \int_{-\infty}^\infty \int_{-\infty}^\infty \calA_m(\tau_i, \bm{p}) G(\Delta \tau_i; x,y,\theta | \bm{p}) d\bm{p} = \sum_{m=1}^3 A_m,
\end{align}
\noindent where $A_m, \ m \in[1,3]$ denotes the corresponding 3D integral, and for $\calA_m$ we have
\begin{align} \label{calA}
\calA_1 &= - \left[2 \varepsilon \bar{\eta} (\xi - x_k) + r\right] \calC(\bm{p}), \quad
\calA_2 = -2 \varepsilon \mu_y(\nu, \bm{p}) (\eta - y_j) \calC(\bm{p}), \quad
\calA_3 = -2 \varepsilon \mu_\theta(\nu, \bm{p}) (\zeta - \theta_l) \calC(\bm{p}).
\end{align}
\noindent where $\mu_x, \mu_y, \mu_\theta$ are defined in \eqref{Marketron_3D}.

An explicit representation of $A_m$ can be obtained as follows. First, note that the following identities hold
\begin{align}
\int_{-\infty }^{\infty } &\frac{e^{-\varepsilon  (\xi - x_k)^2-\frac{(x-\xi )^2}{2 \sigma ^2 \Delta\tau }}}{\sigma  \sqrt{2 \pi \Delta\tau }} \, d\xi = \frac{1}{a(\sigma)}e^{\frac{-\varepsilon  (x-x_k)^2}{a^2(\sigma)}}, \qquad
\int_{-\infty }^{\infty } \frac{e^{-\varepsilon (\eta - y_1)^2-\frac{(y - \eta)^2}{2 \sigma^2_y \Delta\tau }}}{\sigma_y \sqrt{2 \pi  \Delta\tau }} \, d\eta = \frac{1}{a(\sigma_y)} e^{\frac{-\varepsilon  (y-y_1)^2}{a^2(\sigma_y)}}, \\
\int_{-\infty }^{\infty } &\frac{e^{-\varepsilon  (\zeta - \theta_1)^2-\frac{(\theta - \zeta)^2}{2 \sigma^2_\theta \Delta\tau }}}{\sigma_\theta \sqrt{2 \pi \Delta\tau }} \, d\zeta = \frac{1}{a(\sigma_\theta)} e^{\frac{-\varepsilon  (\theta-\theta_1)^2}{a^2(\sigma_\theta)}}, \nonumber
\end{align}
\noindent where $a(\sigma) = \sqrt{1 + 2 \bar{\epsilon} \sigma^2 \Delta\tau}$. Accordingly,
\begin{align} \label{A_coef}
A_3 &=  - 2 \varepsilon \frac{k \alpha_1 }{a^4(\sigma_\theta)} \Omega_{kjl}(x,y,\theta), \quad
\alpha_1 = a^2(\sigma_\theta) \sigma^2_\theta \Delta \tau + (\theta - \theta_l)[ \theta - \theta_l + a^2(\sigma_\theta)(\theta_l - \bar{\theta})], \\
A_1 &= - \left[ 2 \varepsilon \frac{\bar{\eta}(x-x_k)}{a^2(\sigma)} + r \right] \Omega_{kjl}(x,y,\theta), \nonumber \\
\Omega_{kjl}(x,y,\theta) &= \frac{1}{a(\sigma) a(\sigma_y) a(\sigma_\theta)}e^{\frac{-\varepsilon (x-x_k)^2}{a^2(\sigma)}}
 e^{\frac{-\varepsilon (y-y_j)^2}{a^2(\sigma_y)}} e^{\frac{-\varepsilon (\theta-\theta_l)^2}{a^2(\sigma_\theta)}}. \nonumber
\end{align}

\paragraph{Evaluation of $A_{2}$.} From the definitions in \eqref{upsA}, \eqref{calA}, we have
\begin{align} \label{A2}
A_2 &= - 2 \varepsilon \int_{-\infty}^\infty \int_{-\infty}^\infty \int_{-\infty}^\infty \mu_y(\nu, \bm{p}) (\eta - y_j)
\frac{e^{-\varepsilon  (\xi - x_k)^2-\frac{(x-\xi )^2}{2 \sigma^2 \Delta\tau }}}{\sigma  \sqrt{2 \pi  \Delta\tau }}
\frac{e^{-\varepsilon  (\eta - y_j)^2-\frac{(y-\eta )^2}{2 \sigma^2_y \Delta\tau }}}{\sigma_y  \sqrt{2 \pi  \Delta\tau }}
\frac{e^{-\varepsilon  (\zeta - \theta_l)^2-\frac{(\theta-\zeta)^2}{2 \sigma^2_\theta \Delta\tau }}}{\sigma_\theta  \sqrt{2 \pi  \Delta\tau }}
\, d\bm{p} \\
&= - 2 \varepsilon \left[ A_{21} + A_{22} - A_{23} \right]. \nonumber
\end{align}

Using the definition of $\mu_y(\nu, \bm{p})$ from \eqref{Marketron_3D}, we obtain
\begin{equation}
A_{21} = \frac{1}{a(\sigma)}e^{\frac{-\varepsilon(x-x_k)^2}{a^2(\sigma)}}
\frac{y - y_j}{a^3(\sigma_y)}e^{\frac{-\varepsilon(y-y_j)^2}{a^2(\sigma_y)}}
\frac{1}{{\sigma_\theta  \sqrt{2 \pi  \Delta\tau }}}
\int_{-\infty}^\infty h(\zeta) e^{-\varepsilon(\zeta - \theta_l)^2 - \frac{(\theta-\zeta)^2}{2 \sigma^2_\theta \Delta\tau }} d\zeta.
\end{equation}

The last integral in the above expression can be evaluated similarly to $I_{12}$ in \eqref{I12}, replacing $a_1(\tau)$ with $a_2(\tau)$ and $b_1$ with $b_2$. Furthermore,
\begin{align}
A_{22} = \frac{\mu (\tilde{y} - \sigma_y^2 \Delta \tau)}{a^2(\sigma_y)} \Omega_{kjl}(x,y,\theta), \qquad \tilde{y} = (y - y_j)(\bar{y} - y).
\end{align}
Finally,
\begin{align}
A_{23} &= \frac{c}{a(\sigma_\theta)}e^{\frac{-\varepsilon(\theta - \theta_l)^2}{a^2(\sigma_\theta)}}\frac{y - y_j}{a^3(\sigma_y)} e^{\frac{-\varepsilon(y-y_j)^2}{a^2(\sigma_y)}} I_{23}, \quad
I_{23} = \int_{-\infty }^{\infty} V_M(\xi) \frac{e^{-\varepsilon  (\xi - x_k)^2-\frac{(x-\xi )^2}{2 \sigma^2 \Delta\tau }}}{\sigma  \sqrt{2 \pi  \Delta\tau }} d\xi.
\end{align}
Here, $V_M(x)$ should be taken not from its original definition in \eqref{Langevin_potential_exact}, but rather by integrating $V_M'(x)$ with modification provided in \eqref{R2x}, with replacement of $R_1(x)$ with $R_2(x)$. This yields
\begin{align}
V_M(x) &= \frac{1}{2 \epsilon} \Bigg\{ \erfc(b) + e^{-x}\left[\erf(b+x) - 1\right] + e^{b + 1/4} \left[ \erf \left( b + 1/2 \right) - \erf \left(b + x + 1/2 \right) \right] \Bigg\},
\end{align}
\noindent where $b$ is defined in \eqref{defb}. Thus,
\begin{align}
I_{23} &= \frac{1}{2\epsilon} \left[ \calJ_1 + \calJ_2 + \calJ_3 \right], \qquad \calJ_i = \int_{-\infty }^{\infty} J_i(\xi) \frac{e^{-\varepsilon  (\xi - x_k)^2-\frac{(x-\xi )^2}{2 \sigma^2 \Delta\tau }}}{\sigma  \sqrt{2 \pi  \Delta\tau }} d\xi, \\
J_1(x) &=  B_1 - e^{-x}, \qquad B_1 = \erfc(b) + e^{b + 1/4} \erf \left( b + 1/2 \right), \nonumber \\
J_2(x) &= e^{-x} \erf(b+x), \qquad
J_3(x) = - e^{b + 1/4} \erf \left(b + x + 1/2 \right), \nonumber
\end{align}
\noindent and so
\begin{align}
\calJ_1 &= \frac{1}{a(\sigma)}e^{\frac{-\varepsilon(x-x_k)^2}{a^2(\sigma)}} \left[ B_1 - e^{\chi} \right], \quad
\calJ_2 = I_{v,1}, \quad \calJ_3 = - e^{b + 1/4} I_{v,1}\Big|_{\stackrel{b \to b + 1/2}{\beta \to \beta_0}}, \quad \beta_0 = - \frac{x}{a(\sigma) \sigma \sqrt{2\Delta\tau}},
\end{align}
\noindent where $I_{v,1}$ is defined and computed in \cref{evI13}.

\subsection{The source term of function $\calK(\tau_m, x, y, \theta; \calC)$} \label{appSource}

The source term in $\calK(\tau_m, x, y, \theta; \calC)$ originates from the last term in the definition of $\bar{\Phi}(\tau_m, \bm{p}; \calC)$, which is independent of $\calC$, and, consequently, the RBF coefficients. As a result, it is expressed as
\begin{align}
\Upsilon_0(\tau_i, x, y, \theta) &= - \frac{1}{2 \gamma \sigma^2} \int_{-\infty}^\infty \int_{-\infty}^\infty \int_{-\infty}^\infty \bar{\mu}^2_x(\tau_i, \bm{p}) G(\Delta \tau; x,y,\theta| \bm{p}) d\bm{p}, \\
\bar{\mu}_x(\tau_i, \bm{p}) &= f_1(\zeta)  -  c(\tau_i) \eta V'_M(\xi).
\end{align}
This can be re-written in the form
\begin{equation}
\Upsilon_0(\tau_i, x, y, \theta) = - \frac{1}{2 \gamma \sigma^2} \left( \bar{A}_{1} + \bar{A}_{2} + \bar{A}_{3} \right), \\
\end{equation}
\noindent where
\begin{align}
\bar{A}_{1} &= \frac{1}{a(\sigma)} e^{\frac{-\varepsilon(x - x_k)^2}{a^2(\sigma)}} \frac{1}{a(\sigma_y)}e^{\frac{-\varepsilon(y - y_j)^2}{a^2_y(\sigma)}} \bar{A}_{f,2}, \qquad
\bar{A}_{2} = - 2 c(\tau_i) \frac{y - y_j + y_j a^2(\sigma_y)}{a^3(\sigma_y)}e^{\frac{-\varepsilon(y - y_j)^2}{a^2_y(\sigma)}} \bar{A}_{f,1} \bar{A}_{v,1}, \\
\bar{A}_{3} &= c^2(\tau_i) \frac{1}{a(\sigma_\theta)}e^{\frac{-\varepsilon(\zeta - \theta_l)^2}{a^2(\sigma_\theta)}}
\frac{\lambda}{a^5(\sigma_y)}e^{\frac{-\varepsilon(y - y_j)^2}{a^2_y(\sigma)}} \bar{A}_{v,2}, \qquad \lambda = \left[ y - y_j + y_j a^2(\sigma_y) \right]^2 + a^2(\sigma_y) \sigma^2_y \Delta \tau, \nonumber \\
\bar{A}_{f,i} &= \frac{1}{\sigma_\theta \sqrt{2 \pi \Delta\tau }} \int_{-\infty }^{\infty } f^i_1(\zeta) e^{- \frac{(\theta - \zeta)^2}{2 \sigma^2_\theta \Delta\tau }} \, d\zeta, \qquad
\bar{A}_{v,i} = \frac{1}{\sigma\sqrt{2 \pi \Delta\tau}} \int_{-\infty }^{\infty} \left[ V'_M(\xi) \right]^i e^{-\frac{(x-\xi)^2}{\sigma  \sqrt{2 \pi  \Delta\tau}}} d\xi. \nonumber
\end{align}

It follows from \cref{appB} that
\begin{align}
\bar{A}_{f,1} &= I_{f,1}\Big|_{\stackrel{\varepsilon \to 0}{\theta_l \to 0}}, \qquad
\bar{A}_{v,1} = - I_{v,1}\Big|_{\stackrel{\varepsilon \to 0}{\theta_l \to 0}}, \qquad \bar{A}_{f,2} = I_{f,2}\Big|_{\stackrel{\varepsilon \to 0}{\theta_l \to 0}}, \qquad \bar{A}_{v,2} = I_{v,2}\Big|_{\stackrel{\varepsilon \to 0}{\theta_l \to 0}}.
\end{align}

\subsection{The quadratic part of function $\calK(\tau_m, x, y, \theta; \calC)$} \label{appSquare}

Given a set of collocation points $\bm{c}, \bm{c}^*$, the quadratic in $\calC$ component of $\calK(\tau_i, x, y, \theta; \calC)$ can be expressed as
\begin{align} \label{Pi}
Q(\tau_i, x, y, \theta) &= \sum_{k,j,l} \sum_{k^*,j^*,l^*} c_{i,kjl} c_{i,k^*j^*l^*}
\int_{-\infty}^\infty \int_{-\infty}^\infty \int_{-\infty}^\infty \bar{\Psi}_2(\tau_i, \bm{p}, \bm{c}, \bm{c}^*) G(\Delta \tau; x,y,\theta| \bm{p}) d\bm{p},
\end{align}
\noindent where $\bar{\Psi}_2(\tau_i, \bm{p}, \bm{c}, \bm{c}^*)$ is defined in \eqref{defPsi12} as
\begin{equation}
\bar{\Psi}_2(\tau_i, \bm{p}, \bm{c}, \bm{c}^*) = \frac{1}{2} \gamma \sigma_y^2 \calC^2_y(\bm{p},\bm{c}, \bm{c}^*) + \frac{1}{2} \gamma \sigma_\theta^2 \calC^2_\theta(\bm{p}, \bm{c}, \bm{c}^*).
\end{equation}
Accordingly,
\begin{align}
\int_{-\infty}^\infty &\int_{-\infty}^\infty \int_{-\infty}^\infty \bar{\Psi}_2(\tau_i, \bm{p}, \bm{c}, \bm{c}^*) G(\Delta \tau; x,y,\theta| \bm{p}) d\bm{p} = \frac{1}{2} \gamma \sigma_y^2 I_y(\bm{q}, \bm{c}, \bm{c}^*) + \frac{1}{2} \gamma \sigma_\theta^2 I_\theta(\bm{q}, \bm{c}, \bm{c}^*), \\
I_y(\bm{q}, \bm{c}, \bm{c}^*)  &= \int_{-\infty}^\infty \frac{e^{-\varepsilon (\xi - x_k)^2 -\varepsilon (\xi - x_{k^*})^2 - \frac{(x-\xi)^2}{2 \sigma^2 \Delta\tau }}}{\sigma \sqrt{2 \pi  \Delta\tau }} d\xi +
\int_{-\infty}^\infty (\eta - y_j)(\eta - y_{j^*}) \frac{e^{-\varepsilon (\eta - y_j)^2 -\varepsilon (\eta - y_{j^*})^2 - \frac{(y-\eta)^2}{2 \sigma^2_y \Delta\tau }}}{\sigma_y \sqrt{2 \pi  \Delta\tau }} d\eta \nonumber \\
&+ \int_{-\infty}^\infty \frac{e^{-\varepsilon (\zeta - \theta_l)^2 -\varepsilon (\zeta - \theta_{l^*})^2 - \frac{(\theta-\zeta)^2}{2 \sigma^2_\theta \Delta\tau }}}{\sigma_\theta \sqrt{2 \pi  \Delta\tau }} d\zeta, \nonumber \\
I_\theta(\bm{q}, \bm{c}, \bm{c}^*)  &= \int_{-\infty}^\infty \frac{e^{-\varepsilon (\xi - x_k)^2 -\varepsilon (\xi - x_{k^*})^2 - \frac{(x-\xi)^2}{2 \sigma^2 \Delta\tau }}}{\sigma \sqrt{2 \pi  \Delta\tau }} d\xi +
\int_{-\infty}^\infty \frac{e^{-\varepsilon (\eta - y_j)^2 -\varepsilon (\eta - y_{j^*})^2 - \frac{(y-\eta)^2}{2 \sigma^2_y \Delta\tau }}}{\sigma_y \sqrt{2 \pi  \Delta\tau }} d\eta \nonumber \\
&+ \int_{-\infty}^\infty (\zeta - \theta_l)(\zeta - \theta_{l^*})  \frac{e^{-\varepsilon (\zeta - \theta_l)^2 -\varepsilon (\zeta - \theta_{l^*})^2 - \frac{(\theta-\zeta)^2}{2 \sigma^2_\theta \Delta\tau }}}{\sigma_\theta \sqrt{2 \pi  \Delta\tau }} d\zeta. \nonumber
\end{align}
The evaluation of all the aforementioned integrals can be performed in closed form, yielding
\begin{align}
I_y(\bm{q}, \bm{c}, \bm{c}^*) &= \frac{1}{a^4(\sigma_y\sqrt{2})} \Pi(\sigma, x_k, x_{k^*}) \Pi(\sigma_y, y_j, y_{j^*}) \Pi(\sigma_\theta, \theta_l, \theta_{l^*}) \\
 &\cdot \left[ \left(y - \frac{1}{2} (y_j + y_{j^*}) \right)^2 - \frac{a^2(\sigma_y \sqrt{2})}{4} \left( 4 \sigma^2_y \Delta \tau - (y_j - y_{j^*}) a^2(\sqrt{2} \sigma_y) \right) \right], \nonumber \\
I_\theta(\bm{q}, \bm{c}, \bm{c}^*)  &= \frac{1}{a^4(\sigma_\theta\sqrt{2})} \Pi(\sigma, x_k, x_{k^*}) \Pi(\sigma_y, y_j, y_{j^*}) \Pi(\sigma_\theta, \theta_l, \theta_{l^*}) \nonumber \\
&\cdot \left[ \left(\theta - \frac{1}{2} (\theta_l + \theta_{l^*}) \right)^2 - \frac{a^2(\sigma_\theta \sqrt{2})}{4} \left( 4 \sigma^2_\theta \Delta \tau - (\theta_l - \theta_{l^*}) a^2(\sqrt{2} \sigma_\theta) \right) \right], \nonumber \\
\Pi(\sigma, x_k, x_{k^*}) &= \frac{1}{a(\sigma\sqrt{2})} e^{\frac{-\varepsilon [ (x-x_k)^2 + (x-x_{k^*})^2
- 2 \varepsilon \sigma^2 \Delta \tau (x_k - x_{k^*})^2]}{a^2(\sigma\sqrt{2})}}. \nonumber
\end{align}

\section{Calculation of auxiliary integrals} \label{appB}

This section is dedicated to computing four auxiliary expressions $I_{f,1}, I_{f,2}, I_{v,1}, I_{v,2}$ which we need to evaluate $A_2$ in \eqref{A2} and the integrals of the source term in \cref{appSource}. By definition, they read
\begin{align} \label{Iinteg}
I_{f,1} &= \int_{-\infty }^{\infty } f(\zeta) \frac{ e^{-\varepsilon  (\zeta - \theta_l)^2 - \frac{(\theta - \zeta)^2}{2 \sigma^2_\theta \Delta\tau }}}{\sigma_\theta  \sqrt{2 \pi  \Delta\tau }} \, d\zeta, \qquad
I_{v,1} = -\int_{-\infty }^{\infty } \left( 1 - \frac{g}{e^{\xi} + \bar{\epsilon} g} \right) \frac{e^{-\varepsilon  (\xi - x_k)^2-\frac{(x-\xi )^2}{2 \sigma ^2 \Delta\tau } - \xi}}{\sigma  \sqrt{2 \pi  \Delta\tau }} \, d\xi, \\
I_{f,2} &= \int_{-\infty }^{\infty } f^2(\zeta) \frac{ e^{-\varepsilon  (\zeta - \theta_l)^2 - \frac{(\theta - \zeta)^2}{2 \sigma^2_\theta \Delta\tau }}}{\sigma_\theta  \sqrt{2 \pi  \Delta\tau }} \, d\zeta, \qquad
I_{v,2} = \int_{-\infty }^{\infty } \left( 1 - \frac{g}{e^{\xi} + \bar{\epsilon} g} \right)^2 \frac{e^{-\varepsilon  (\xi - x_k)^2-\frac{(x-\xi )^2}{2 \sigma ^2 \Delta\tau } - 2\xi}}{\sigma  \sqrt{2 \pi  \Delta\tau }} \, d\xi. \nonumber
\end{align}

\subsection{Evaluation of $I_{v,1}$.} \label{evI13}

\begin{figure}[!htb]
\centering
\includegraphics[width=0.52\textwidth]{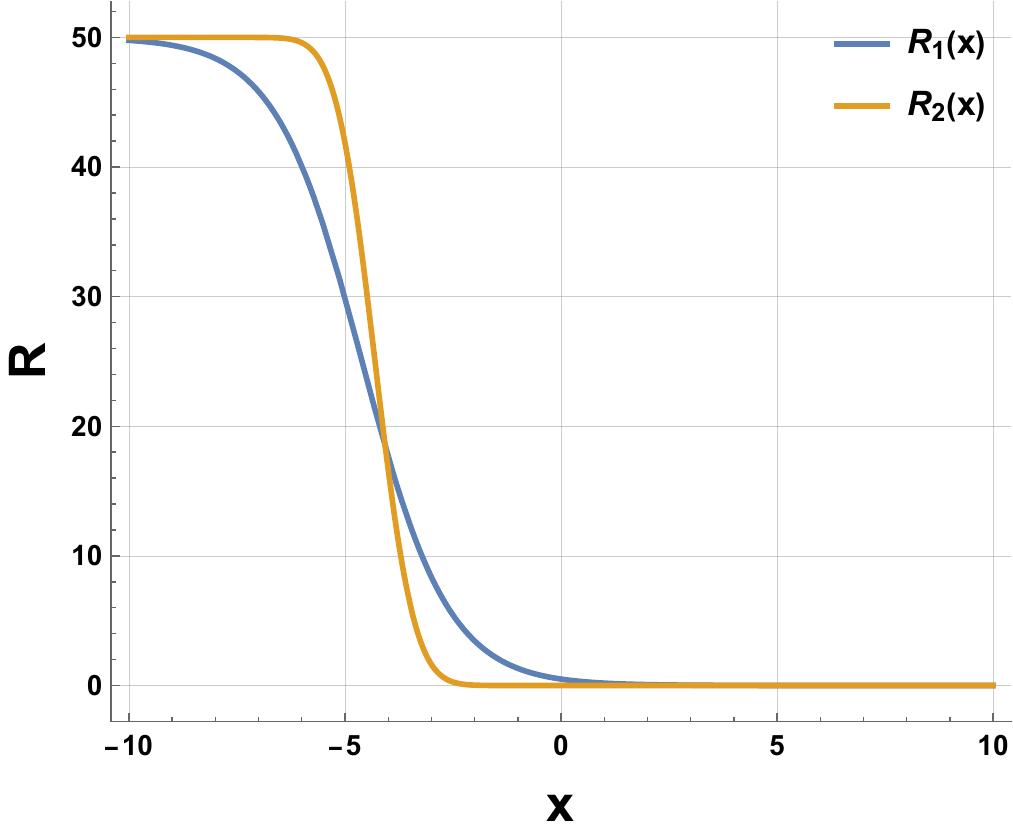}
\caption{Functions $R_1(x), R_2(x)$ at $\bar{\epsilon} = 0.1, g = 0.5$.}
\label{regul}
\end{figure}

Unfortunately, this integral cannot be taken in closed form. To resolve this, we slightly change the definition of $V'_M(x)$ in \eqref{Langevin_potential_exact}. Indeed, the expression $1 - R_1(x)$ with $R_1(x) = \frac{g}{e^{x} + \bar{\epsilon} g}$ was introduced in \cite{HalperinItkin2025ar} as a regularization term to prevent divergence of the SDE at $x \to -\infty$ because when $x$ changes from $+\infty$ to $-\infty$, this term changes sign (thus, providing a mean-reversion in the drift). However, the same effect can be achieved by using another regularizer which is similar to $R_1$, for instance
\begin{equation} \label{R2x}
R_2(x) = \frac{1}{2 \bar{\epsilon}} \left[ 1 - \erf(x + b) \right],
\end{equation}
\noindent where $b$ could be chosen, e.g., as
\begin{equation} \label{defb}
b = 4 - \erf^{-1} \left[1 - \frac{2}{1 + e^4 g \bar{\epsilon} } \right].
\end{equation}
In Fig.~\ref{regul}, the corresponding plots of $R_1(x)$ and $R_2(x)$ are presented for $\bar{\epsilon} = 0.1, g = 0.5$. It can be seen that these functions look similar, therefore either of them can be used as a suitable regularizer.

Nevertheless, as shown in the following Proposition, replacing $R_1(x)$ with $R_2(x)$ offers the advantage that $I_{v,1}$ can be computed in closed form.
\begin{proposition} \label{prop1}
The following identity holds
\begin{align} \label{prop1Form}
I_{v,1} &= -\left( 1 - \frac{1}{2\bar{\epsilon}} \right) I_1 - \frac{1}{2\bar{\epsilon}} I_{2}, \\
I_1 &= \frac{1}{a(\sigma)}  e^{\frac{-\varepsilon (x - x_k)^2}{a^2(\sigma)}} e^\chi, \qquad
I_{2} = \frac{e^{\chi}}{\sigma \sqrt{2 \pi  \Delta\tau }} \frac{\sqrt{\pi}}{\alpha} \erf \left( \frac{\alpha b - \beta}{\sqrt{1 + \alpha^2}} \right), \nonumber \\
\alpha &= \frac{a(\sigma)}{\sigma \sqrt{2 \Delta\tau}}, \qquad \beta = \frac{(x_k - x) - x_k a^2(\sigma) + \sigma^2 \Delta\tau}{a(\sigma) \sigma \sqrt{2\Delta\tau}}, \qquad
\chi = \frac{- \varepsilon(x - x_k)^2 + \sigma^2 \Delta\tau/2}{a^2(\sigma)}. \nonumber.
\end{align}

\begin{proof}[{\bf Proof}]
We need to evaluate the integral
\begin{equation}
I_{v,1} = -\int_{-\infty }^{\infty } \left[ 1 - \frac{1}{2 \bar{\epsilon}}  \left( 1 - \erf(\xi + b) \right) \right] \frac{e^{-\varepsilon  (\xi - x_k)^2-\frac{(x-\xi )^2}{2 \sigma ^2 \Delta\tau } - \xi}}{\sigma  \sqrt{2 \pi  \Delta\tau }} \, d\xi = -\left(1 - \frac{1}{2\bar{\epsilon}}\right) I_1 - \frac{1}{2\bar{\epsilon}} I_{2}.
\end{equation}
\noindent where $I_1$ and $I_{2}$ correspond to the terms with a constant and $\erf(\xi + b)$, respectively. The first integral can be computed in closed form to yield
\begin{align}
I_1 &= \frac{1}{a(\sigma)}  e^{\frac{-\varepsilon (x - x_k)^2}{a^2(\sigma)}} e^\chi.
\end{align}

In the second integral $I_{2}$, we complete the squares to obtain
\begin{gather} \label{I22fin}
I_{2} = \frac{e^{\chi}}{\sigma \sqrt{2 \pi  \Delta\tau }} \int_{-\infty}^\infty \erf(\xi + b) e^{ -(\alpha \xi + \beta)^2} d\xi.
\end{gather}
The integral in $I_{2}$ is a known result (see, e.g., \cite{Briggs2003}) and is given by
\begin{equation} \label{I22}
\int_{-\infty}^\infty \erf(\xi + b) e^{ -(\alpha \xi + \beta)^2} d\xi = \frac{\sqrt{\pi}}{\alpha} \erf \left( \frac{\alpha b - \beta}{\sqrt{1 + \alpha^2}} \right).
\end{equation}
Substituting this expression into \eqref{I22fin} completes the proof.
\end{proof}
\end{proposition}

\subsection{Evaluation of $I_{f,1}$} \label{evI12}

\begin{figure}[!htb]
\centering
\includegraphics[width=0.52\textwidth]{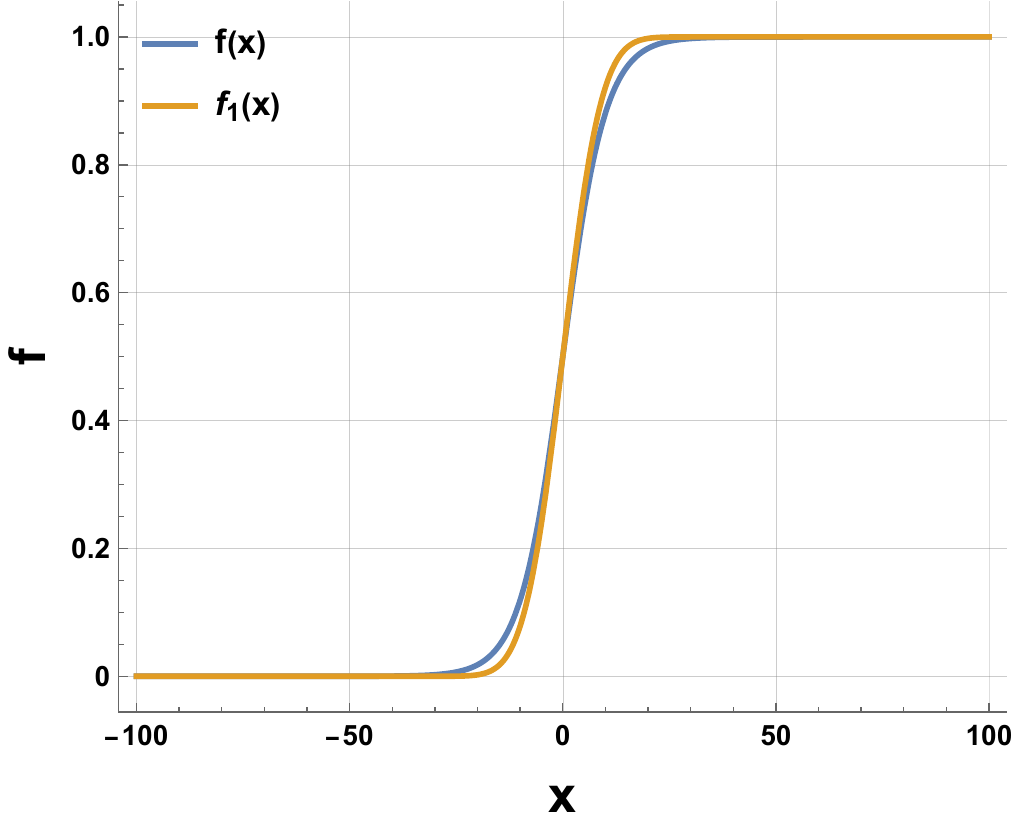}
\caption{Functions $f(x), f_1(x)$ at $b_1 = 0.2$.}
\label{f_theta}
\end{figure}

By definition of $f(\theta)$ in \eqref{fhparam}, we have
\begin{align}
I_{f,1} &= \frac{1}{\sigma_\theta  \sqrt{2 \pi  \Delta\tau }} \int_{-\infty }^{\infty } f(\zeta) e^{-\varepsilon (\zeta - \theta_l)^2 - \frac{(\theta - \zeta)^2}{2 \sigma^2_\theta \Delta\tau }} d\zeta = \frac{1}{\sigma_\theta  \sqrt{2 \pi  \Delta\tau }} \int_{-\infty }^{\infty} \frac{a_1(\tau_i)}{1 + e^{-b_1 \zeta}} e^{-\varepsilon  (\zeta - \theta_l)^2 - \frac{(\theta - \zeta)^2}{2 \sigma^2_\theta \Delta\tau }} \, d\zeta.
\end{align}
Since the above integral cannot be evaluated in closed form, we introduce a modified version of $f(\theta)$, denoted as $f_1(\theta)$
\begin{equation}
f_1(\theta) = a_1(\tau_i) \Psi(b_1 x/\sqrt{2}) = \frac{1}{2}a_1(\tau_i)\left[ 1 + \erf\left( \frac{b_1}{2} \theta \right) \right].
\end{equation}
Fig.~\ref{f_theta} compares $f(x)$ and $f_1(x)$ for $b_1 = 0.2$. The similar behavior of these functions justifies the substitution of $f_1(x)$ for $f(x)$ in our further analysis.

Using the same approach as employed for computing $I_{2}$ in \eqref{I22}, with this substitution we obtain
\begin{align} \label{I12}
I_{f,1} &= \frac{a_1(\tau_i)}{2} \int_{-\infty }^{\infty} \left[ 1 + \erf\left( \frac{b_1}{2} \zeta \right) \right] \frac{ e^{-\varepsilon  (\zeta - \theta_l)^2 - \frac{(\theta - \zeta)^2}{2 \sigma^2_\theta \Delta\tau}}}{\sigma_\theta  \sqrt{2 \pi  \Delta\tau}} \, d\zeta \\
&= \frac{a_1(\tau_i)}{2 \sigma_\theta}  \left[ e^{\frac{-\varepsilon (\theta-\theta_l)^2}{a^2(\sigma_\theta)}}  +
\frac{e^{\chi_\theta}}{\sqrt{2 \pi  \Delta\tau}} \frac{\sqrt{\pi}}{\alpha_\theta} \erf \left( \frac{d\beta_\theta}{\sqrt{d^2 + \alpha^2_\theta}} \right) \right], \nonumber \\
\beta_\theta &= \frac{(\theta_l - \theta) - \theta_l a^2(\sigma_\theta)}{a(\sigma_\theta) \sigma_\theta \sqrt{2\Delta\tau}}, \quad \chi_\theta = -\frac{\varepsilon(\theta - \theta_l)^2}{a^2(\sigma_\theta)}, \quad d = \frac{b_1}{2}, \quad \alpha_\theta = \frac{a(\sigma_\theta)}{\sigma_\theta \sqrt{2 \Delta\tau}}. \nonumber
\end{align}

\subsection{Evaluation of $I_{f,2}$} \label{evIf2}

By definition of $f(\theta)$ in \eqref{fhparam}, we have
\begin{align}
I_{f,2} &= \frac{a^2_1(\tau_i)}{4} \int_{-\infty }^{\infty} \left[ 1 + \erf\left( \frac{b_1}{2} \zeta\right) \right]^2 \frac{ e^{-\varepsilon  (\zeta - \theta_l)^2 - \frac{(\theta - \zeta)^2}{2 \sigma^2_\theta \Delta\tau }}}{\sigma_\theta  \sqrt{2 \pi  \Delta\tau }} \, d\zeta = \frac{a^2_1(\tau_i)}{4 \sigma_\theta}  \Big[ e^{\frac{-\varepsilon (\theta-\theta_l)^2}{a^2(\sigma_\theta)}}  \\
&+ 2\frac{e^{\chi_\theta}}{\sqrt{2 \pi  \Delta\tau}} \frac{\sqrt{\pi}}{\alpha_\theta} \erf \left( \frac{d\beta_\theta}{\sqrt{d^2 + \alpha^2_\theta}} \right) + \sigma_\theta I_{22} \Big], \qquad
I_{22} = \int_{-\infty }^{\infty} \erf^2\left( \frac{b_1}{2} \zeta \right) \frac{ e^{-\varepsilon  (\zeta - \theta_l)^2 - \frac{(\theta - \zeta)^2}{2 \sigma^2_\theta \Delta\tau }}}{\sigma_\theta  \sqrt{2 \pi  \Delta\tau }} \, d\zeta. \nonumber
\end{align}

To compute $I_{22}$ we use the identity, \cite{as64}
\begin{equation} \label{erf2}
\erf^2(x) = - \frac{4}{\sqrt{\pi}} \int e^{-x^2} \erf(x) dx,
\end{equation}
\noindent so, $I_{22}$ can be represented as
\begin{align}
I_{22} &=  - \frac{4}{\sqrt{\pi}} \int_{-\infty }^{\infty} \left[  \int e^{- \frac{b^2_1}{4} \zeta^2} \erf \left( \frac{b_1}{2}\zeta \right) d\zeta \right] \frac{ e^{-\varepsilon  (\zeta - \theta_l)^2 - \frac{(\theta - \zeta)^2}{2 \sigma^2_\theta \Delta\tau }}}{\sigma_\theta  \sqrt{2 \pi  \Delta\tau }} \, d\zeta \\
&= \frac{4}{\sqrt{\pi}} \left[ \fp{I_{20}}{\theta} +  \fp{I_{20}}{\theta_l} \right], \qquad
I_{20} = \frac{1}{\sigma_\theta  \sqrt{2 \pi  \Delta\tau }} \int_{-\infty }^{\infty} \erf \left( \frac{b_1}{2}\zeta \right) e^{-\varepsilon  (\zeta - \theta_l)^2 - \frac{(\theta - \zeta)^2}{2 \sigma^2_\theta \Delta\tau } - \frac{b^2_1}{4} \zeta^2} \, d\zeta. \nonumber
\end{align}
The integral $I_{20}$ can be evaluated in the same way as the integral $I_2$ in \eqref{I22fin} by completing the squares and using \eqref{I22}. This yields
\begin{align}
I_{20} &= \frac{e^{\chi_{\theta,1}}}{\sigma_\theta  \sqrt{2 \pi  \Delta\tau }} \frac{\sqrt{\pi}}{\alpha_{\theta,1}} \erf \left( \frac{d\beta_{\theta,1}}{\sqrt{d^2 + \alpha^2_{\theta,1}}} \right), \qquad a_+(\sigma_\theta) = \sqrt{1 + 2(1+\varepsilon)\sigma^2_\theta \Delta \tau}, \\
\alpha_{\theta,1} &= \frac{a_+(\sigma_\theta)}{\sigma_\theta \sqrt{2 \Delta\tau}}, \quad
\beta_{\theta,1} = \frac{(\theta_l - \theta) - \theta_l a^2(\sigma_\theta)}{a_+(\sigma_\theta) \sigma_\theta \sqrt{2\Delta\tau}}, \quad
\chi_{\theta,1} = \frac{ - \varepsilon (\theta - \theta_l)^2 - \theta^2_l a^2(\sigma_\theta) + \theta^2_l - \theta^2}{a^{+2}(\sigma_\theta)}. \nonumber
\end{align}

\subsection{Evaluation of $I_{v,2}$} \label{evIv2}

By \eqref{Iinteg} and the analysis of \cref{evI13}, we need to evaluate the integral
\begin{align} \label{Iv2}
I_{v,2} &= \int_{-\infty }^{\infty } \left( 1 - R_2(\xi) \right)^2 \frac{e^{-\varepsilon  (\xi - x_k)^2-\frac{(x-\xi )^2}{2 \sigma ^2 \Delta\tau } - 2\xi}}{\sigma  \sqrt{2 \pi  \Delta\tau }} \, d\xi =
\left( 1 - \frac{1}{2\bar{\epsilon}} \right) I_1 + \left( 1 - \frac{1}{2\bar{\epsilon}} \right)\frac{1}{\bar{\epsilon}}I_2 + I_3, \\
I_3 &= \frac{1}{4 \epsilon^2} \int_{-\infty }^{\infty } \erf^2\left(\xi + b\right)) \frac{e^{-\varepsilon  (\xi - x_k)^2-\frac{(x-\xi )^2}{2 \sigma ^2 \Delta\tau } - 2\xi}}{\sigma  \sqrt{2 \pi  \Delta\tau }}. \nonumber
\end{align}
Again, using \eqref{erf2}, this can be represented as
\begin{align}
I_3 &= - \frac{1}{\sqrt{\pi} \epsilon^2} \int_{-\infty }^{\infty }  \left( \int e^{-(\xi+b)^2} \erf(\xi+b) d\xi \right)\frac{e^{-\varepsilon  (\xi - x_k)^2-\frac{(x-\xi )^2}{2 \sigma ^2 \Delta\tau } - 2\xi}}{\sigma  \sqrt{2 \pi  \Delta\tau }} d\xi \\
&= - \frac{1}{\pi \epsilon^2 \sigma  \sqrt{2 \Delta\tau}} \int_{-\infty }^{\infty }  \erf(\xi+b) e^{-\varepsilon  (\xi - x_k)^2-\frac{(x-\xi )^2}{2 \sigma ^2 \Delta\tau } - 2\xi -(\xi+b)^2 } d\xi. \nonumber
\end{align}

Completing the squares and using \eqref{I22} yields
\begin{align}
I_3 &= - \frac{e^{\chi_2}}{\sqrt{\pi} \epsilon^2 \alpha_2 \sigma \sqrt{2 \Delta\tau}} \erf \left( \frac{\alpha_2 b - \beta_2}{\sqrt{1 + \alpha^2_2}} \right), \\
\alpha_{2} &= \frac{a_+(\sigma)}{\sigma\sqrt{2 \Delta\tau}}, \quad
\beta_{2} = \frac{(x_k - x) - x_k a^2(\sigma) + 2 \sigma^2 \Delta\tau (1+b)}{a_+(\sigma) \sigma\sqrt{2\Delta\tau}},
\quad a_-(\sigma) = a(\sigma) - 1, \nonumber \\
\chi_{2} &= \frac{2(1+2b)\sigma^2 \Delta\tau -\varepsilon (x - x_k)^2 - 2(x + x_k a^2_-(\sigma)) - (b+x)^2 - (b+x_k)^2 a^2_-(\sigma)}{a_+(\sigma) \sigma\sqrt{2\Delta\tau}}. \nonumber
\end{align}


\end{document}